\documentclass{aa}
\usepackage{graphicx}
\usepackage{adjustbox}
\usepackage{natbib}
\usepackage{enumerate}
\usepackage{enumitem}
\usepackage{float}
\usepackage{siunitx}
\usepackage{txfonts}
\begin{document}
%

 \title{Searching for g modes: Part II. Unconfirmed g-mode detection in the power spectrum of the time series of round-trip travel time.}

   \author{T.~Appourchaux
          \inst{1} \&
          T.~Corbard\inst{2}
          }

   \institute{Universit\'e Paris-Sud, Institut d'Astrophysique Spatiale, UMR 8617, CNRS, B\^atiment 121, 91405 Orsay Cedex, France
   \and
   Universit\'e C\^ote d'Azur, Observatoire de la C\^ote d'Azur, CNRS, Laboratoire Lagrange, CS 34229, Nice Cedex 4, France
          }

   \date{Accepted March 4, 2019}

 
  \abstract
  {The recent claims of g-mode detection have restarted the search for these potentially extremely important modes.  The claimed detection of g modes was obtained from the analysis of the power spectrum of the time series of round-trip travel time of p modes.}
   {The goal of this paper is to reproduce these results on which the claims are based for confirming or invalidating the detection of g modes with the method used to make the claims.}
   {We computed the time series of round-trip travel time using the procedure given in Fossat et al. (2017), and used different variations of the times series for comparison.  We used the recently calibrated GOLF data (published in Paper I) with different sampling, different photomultipliers, different length of data for reproducing the analysis.  We also correlated the power spectrum with an asymptotic model of g-mode frequencies in a similar manner to Fossat and Schmider (2018).  We devised a scheme for optimising the correlation both for pure noise and for the GOLF data.}
   {We confirm the analysis performed in Fossat et al. (2017) but draw different conclusions.  Their claims of detection of g modes cannot be confirmed when changing parameters such as sampling interval, length of time series, or photomultipliers.  Other instrument such as GONG and BiSON do not confirm their detection.  We also confirm the analysis performed in Fossat and Schmider (2018), but again draw different conclusions.  For GOLF, the correlation of the power spectrum with the asymptotic model of g-mode frequencies for $l=1$ and $l=2$ show a high correlation at lag=0 and at lag corresponding to the rotational splitting $\nu_l$, but the same occurs for pure noise due to the large number of peaks present in the model.  In addition, other very different parameters defining the asymptotic model also provide a high correlation at these lags.  We conclude that the detection performed in Fossat and Schmider (2018) is an artefact of the methodology.}
   {}
\titlerunning{Searching for g modes: Part II: Unconfirmed g-mode detection} 
   \maketitle
%
\section{Introduction}
The detection of g modes remains a major quest of helioseismology.  The benefit of detecting these modes would be to obtain the structure and dynamics of the very inner
core of the Sun.  There have been several claims of g-mode detection \citep[see][for a review]{Appourchaux2010}.  Since this review, the detection claims of \citet{Garcia07} using the asymptotic properties of g-mode periods were not confirmed by \citet{AMB2010} who used a Bayesian approach for detecting the g modes on multiple time series.  

\citet{Fossat2017} (hereafter F17), using the propagation time of the p-mode wave packet, claimed to have detected the signature of g modes.   Very recently, \citet{Schunker2018} showed that using different fitting procedures or using different start times, or cadence different than 4 hours or different smoothing, the prominent peaks at 210 nHz and its acolytes would smear out or even disappear.  In \citet[][]{TA2018}, we also showed that the use of different sampling time (i.e. 60 s instead of 80 s) would also affect the detection level of the peaks shown in Fig.~10 of F17, which is the basis for the g-mode detection claim.

\citet{Fossat2018} (hereafter FS18) pushed  the analysis done in F17 further by matching an asymptotic model of g-mode periods to the original spectrum obtained in F17.  They extended their claims of g-mode detection to spherical harmonic degrees of $l=3$ and $l=4$.  The detection of higher degrees was a surprise because these modes are believed to be even more damped through the convection zone than modes of lower degrees \citep[][]{Appourchaux2010}.  On the other hand, the observables used for the detection in FS18 are not sensitive to displacements on the solar surface, but to displacements below the convection zone.  The reason for the modulation of the convection zone by the g modes is believed to be related to displacement of the bottom part of the convection zone
induced by the g modes.

In order to test the detection claims of F17, we made longer data sets using a new calibration strategy for the GOLF data \citep[][{hereafter Paper I}]{TA2018} in which we revisited the calibration of the GOLF instrument.   The present F17s the second part, which aims to test the solar g-mode detection of F17 and FS18 with our newly calibrated GOLF time series. 

This paper is organised as follows. In Sect.1 we study the g-mode detection claim of F17 using the new calibrated GOLF time series for the same observation duration but for a different sampling time, different photomultipliers, and different sub-series; and also for a longer time series of 22 years.  We also used other helioseismic time series for testing the claimed detection.  In Sect. 2 we study the g-mode detection claim of FS18 using the same GOLF times series as in their F17.  In Sect. 3 we then finally discuss the data and draw our conclusions.


\section{Reproducing the analysis of F17}
In this section, we focus on comparing the results obtained in F17 with their Figs.~10 and 16.  These two figures focus our attention because this is the main basis for their g-mode detection claim.  The remaining part of the analysis done in F17 has been done more deeply in their FS18, which is the subject of the next section.  The procedure used in this paper for reproducing Figs~10 and 16 of F17 is the same as in their paper apart from the smoothing that needed to be corrected as noted by \citet{Schunker2018}.  For reproducing, Fig. 16 of F17, we also used the p-mode mean rotational splitting $\Omega_p$=0.432 $\mu$Hz as in F17.  For each figure (10 or 16), we deduce the rms fluctuations ($\sigma$).  For the autocorrelation (Fig. 10), the number of independent bins $n_{\rm ind}$ is given by the number of bins present in the window of 3.5 $\mu$Hz width divided by 6 (since the power spectrum is smoothed over 6 bins).  The number of independent bins in the sum of the correlation (Fig. 16) is given by the number of g-mode rotation bins (2000) divided by 6 (since the power spectrum is smoothed over 6 bins).   For either diagram, the statistics are Gaussian.  Therefore the 10\% level $x_{\rm cut}$ in units of $\sigma$ is given by solving the following equation:
\begin{equation}
\frac{0.1}{n_{\rm ind}} \approx \frac{1}{\sqrt{2 \pi}}\int_{x_{\rm cut}}^{\infty} {\rm e}^{-\frac{x^2}{2}} {\rm d}x
\end{equation}
where $x$ is the normalised random variable of mean 0 and of rms 1 \citep[See also][]{TA2000}.  The 10\% level corresponds to a probability of the null hypothesis (${\rm H}_0$) to be true of at least 38\% of the time \citep[See][]{Sellke2001}.

The procedure was applied to the new GOLF time series for two observation duration: 16.5 years as in F17 and the longest time series of 22 years obtained in Paper I.  For the 16.5-year time series, we check the detection with: different sampling time (20~s, 60~s, 80~s), and different photomultipliers (PM1, PM2).  For the longest time series sampled at 80 sec, we also used two different sub-series of 11 years.

\begin{figure*}[!]
\centerline{
\vbox{
\hbox{
\includegraphics[width=6 cm,angle=90]{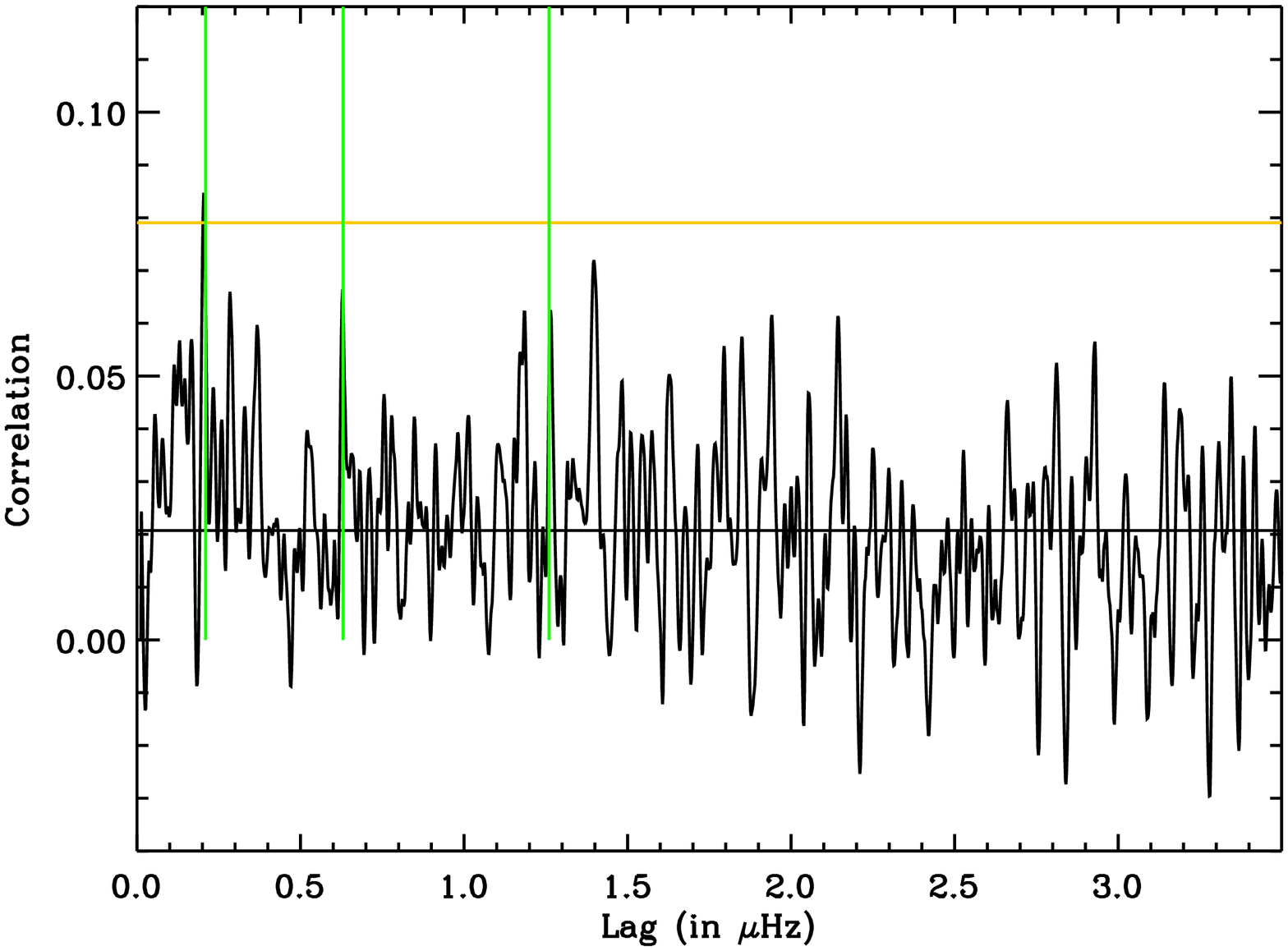}
\includegraphics[width=6 cm,angle=90]{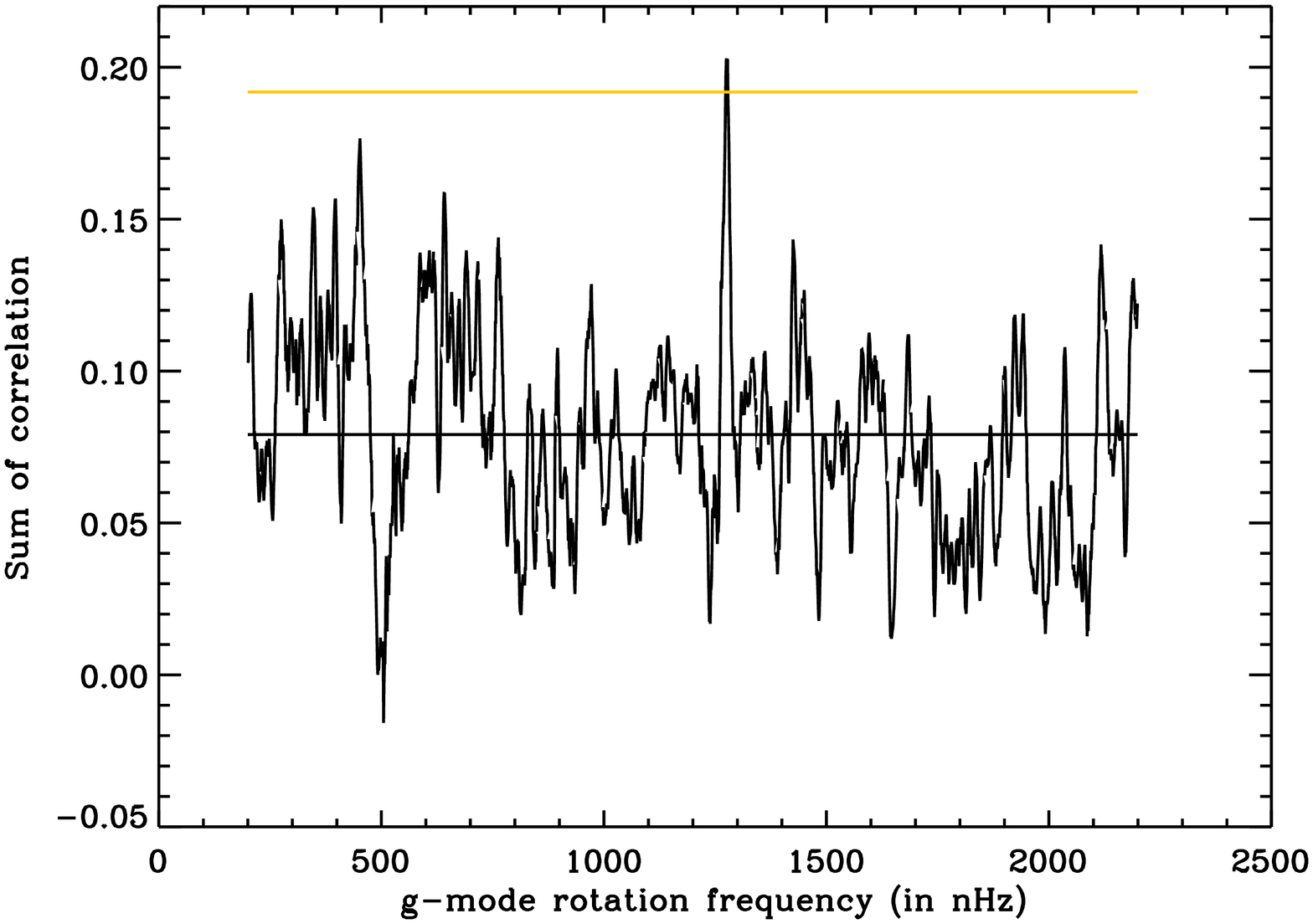}}
\hbox{
\includegraphics[width=6 cm,angle=90]{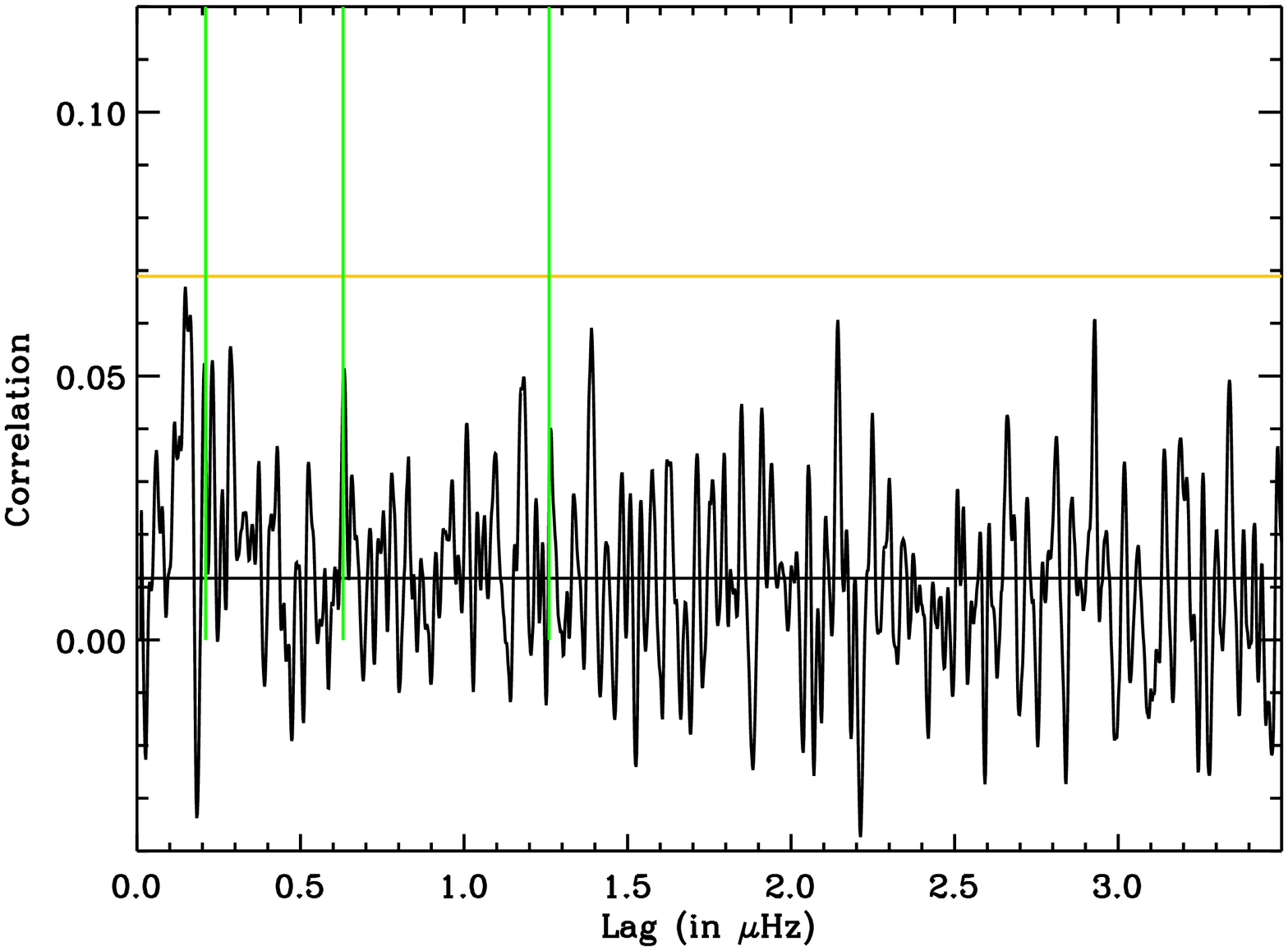}
\includegraphics[width=6 cm,angle=90]{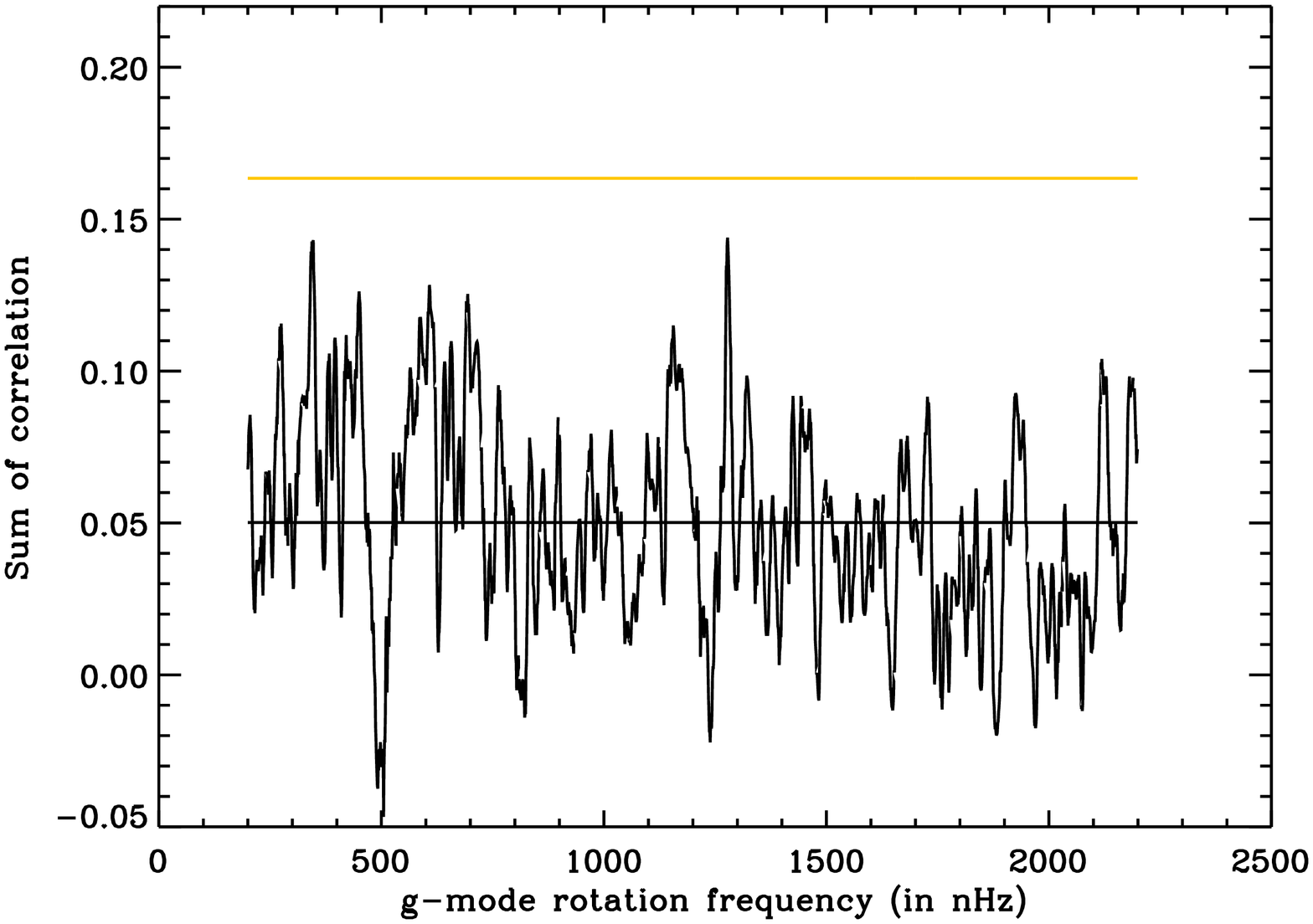}}
\hbox{
\includegraphics[width=6 cm,angle=90]{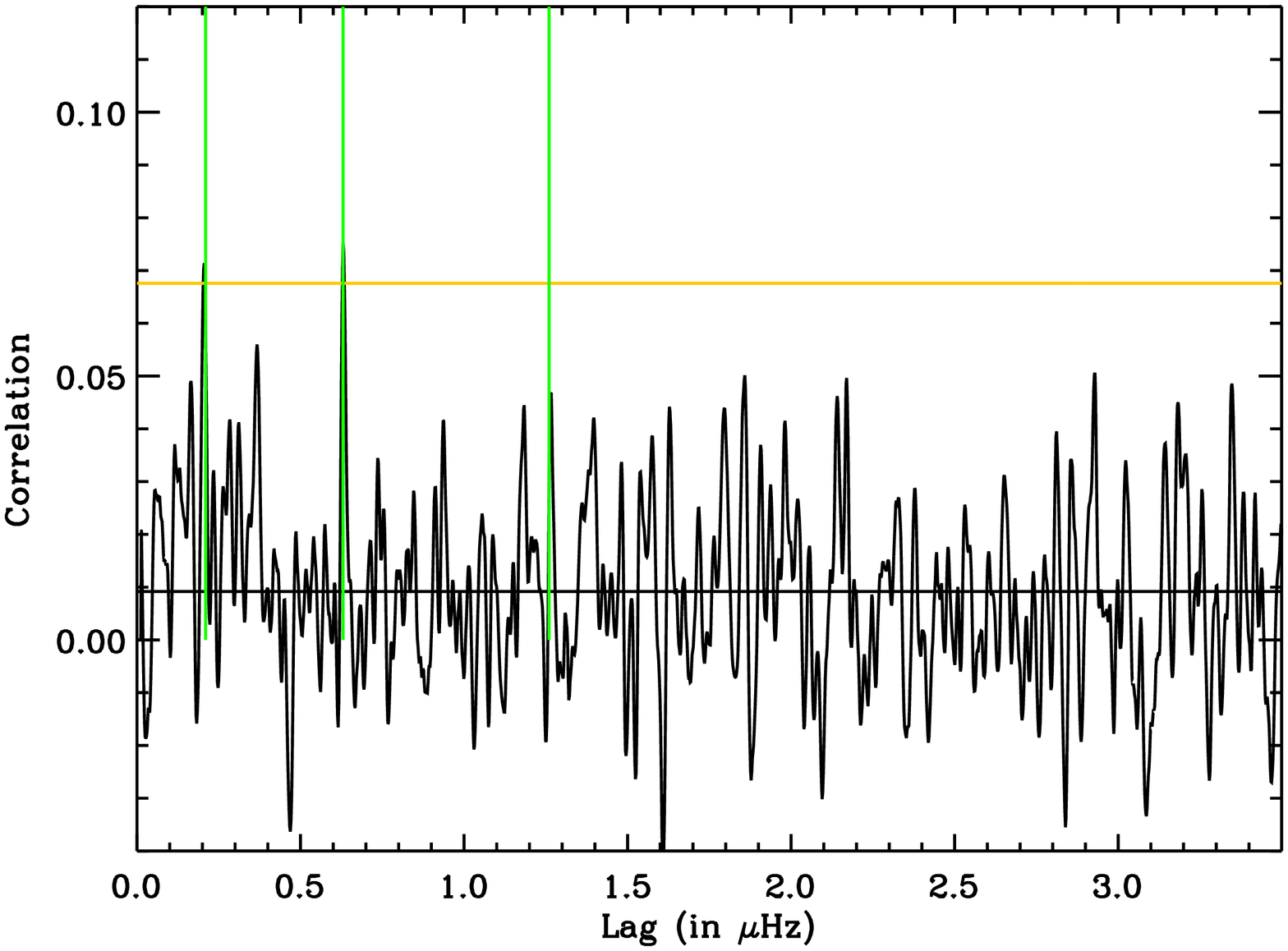}
\includegraphics[width=6 cm,angle=90]{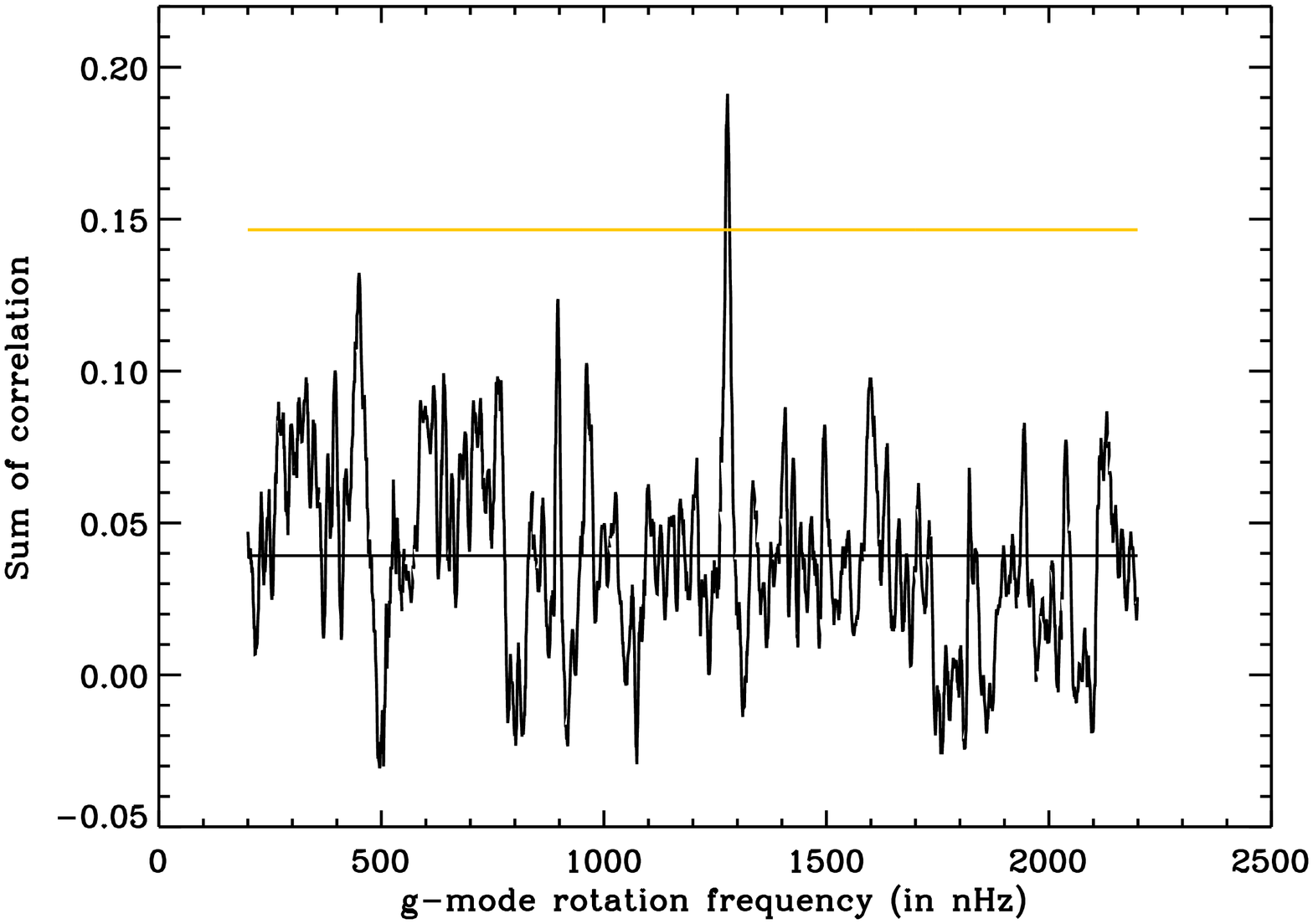}}
}}
\caption{(left) Autocorrelation of the power spectrum as obtained by \citet{Fossat2017} as a function of frequency lag for three different time series of the average of photomultipliers PM1 and PM2 of GOLF sampled at 20 s (top), 60 s (middle) and 80 s (bottom); the bottom panel is comparable to Fig.~10 of \citet{Fossat2017}.  The green vertical lines correspond to frequencies of 210 nHz, 630 nHz and 1260 nHz.  The black line indicates the mean value of the autocorrelation, while the orange line indicates the 10\% probability level that there is at least one peak due to noise in the window; the level is 3.41 $\sigma$.  (right) Sum of the correlation for $l=1$, and $l=2$ modes as obtained by \citet{Fossat2017} as a function of rotation frequency for three different time series sampled at 20 s (top), 60 s (middle) and 80 s (bottom); the bottom panel is comparable to Fig.~16 of \citet{Fossat2017}.  The black line indicates the mean value, while the orange line indicates the 10\% probability level that there is at least one peak in the window due to noise; the level is 3.43 $\sigma$.}
\label{sampling}
\end{figure*}

\subsection{Sampling time}
The original GOLF time series is sampled at 10 s.  This short cadence allowed the compensation for the timing errors of the GOLF instrument, induced by the loss of the On Board Time as shown in Paper I.  From this original cadence, we generated a calibrated cadence of 20 s, from which sampling at 60 s and 80 s can easily be deduced.  Neither F17 nor \citet{Schunker2018} studied the impact of the different sampling time upon the g-mode detection claim.  Given the fact that the travel time is computed with a sampling of 4 h, the original cadence of the data should not have any influence on the detection.  Here we used the newly calibrated GOLF data averaged over the two photomultipliers with the same observation duration as in Paper~I.  Figure~\ref{sampling} shows the results for three different sampling times.  

We can see that the mean correlation drops with the square root of the sampling time.  This is due to the fact that the noise per sample also drops with the same scaling.  On the other hand, the rms of the fluctuations is the same for all three sampling because it does not depend upon the sampling time but only on the observation time.  The peak at 210 nHz in the correlation is above the 10\% threshold for the 20-s and 80-s sampling.  The peak in the correlation at 630 nHz is above the 10\% line only for the 80-s sampling.  Here we note that the peaks at 210 nHz and 630 nHz are both above the threshold level while in F17 only the peak at 210 nHz was.  In the sum of the correlation, we also note that the peak at 1280 nHz is higher than in F17.  Finally, we note that there is no detection in the 60-s sampling.  The detection made in F17 is then confirmed for the 80-s sampling but we cannot exclude that noise plays a significant role in making the peaks in the correlation and the sum of the correlation appear or disappear.  This conclusion is not contradicted by the fact that the null hypothesis is rejected at the 38\% level with the threshold level of 10\%.

\subsection{Different photomultipliers}
\begin{figure*}[!]
\centerline{
\vbox{
\hbox{
\includegraphics[width=6 cm,angle=90]{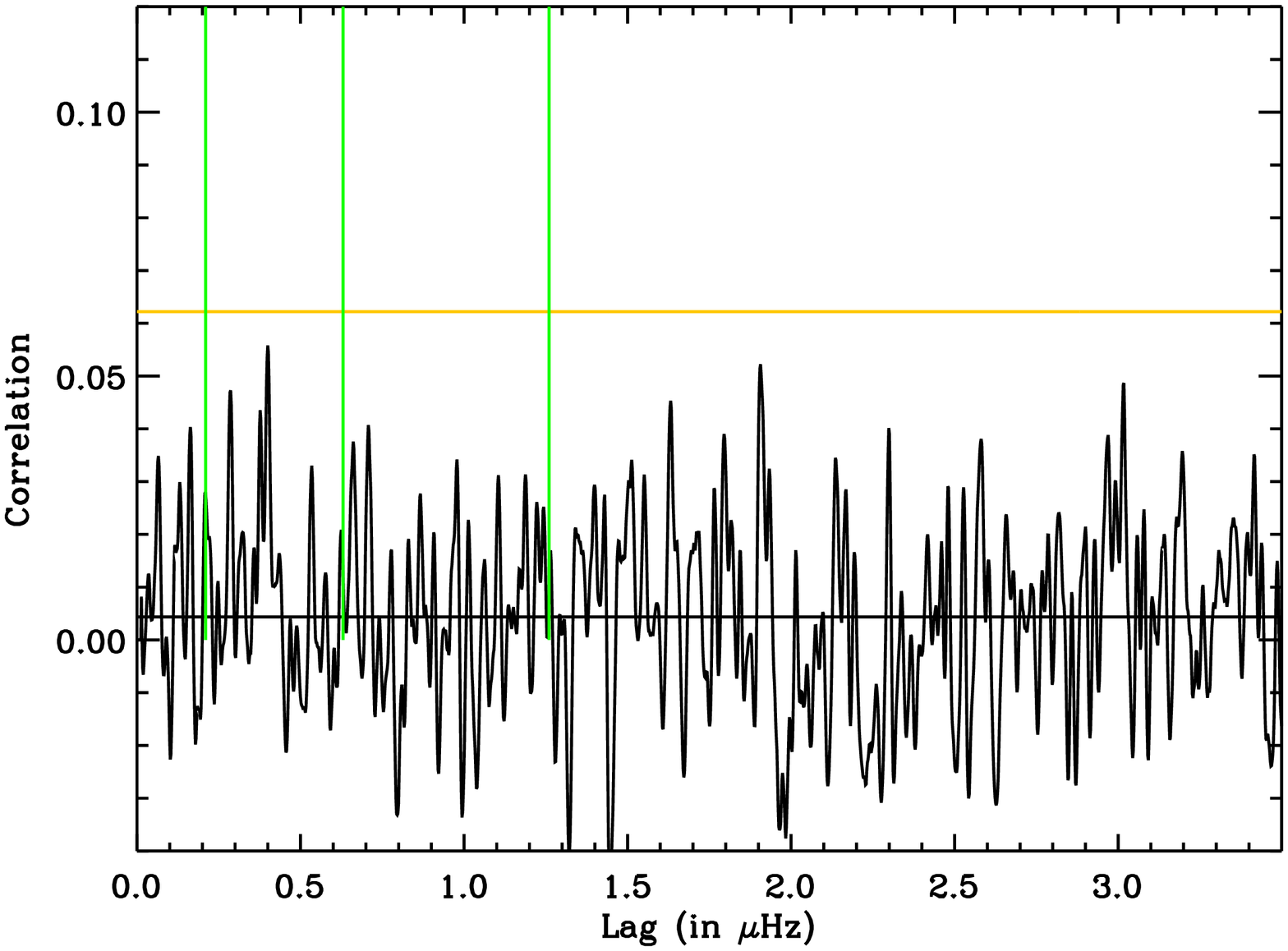}
\includegraphics[width=6 cm,angle=90]{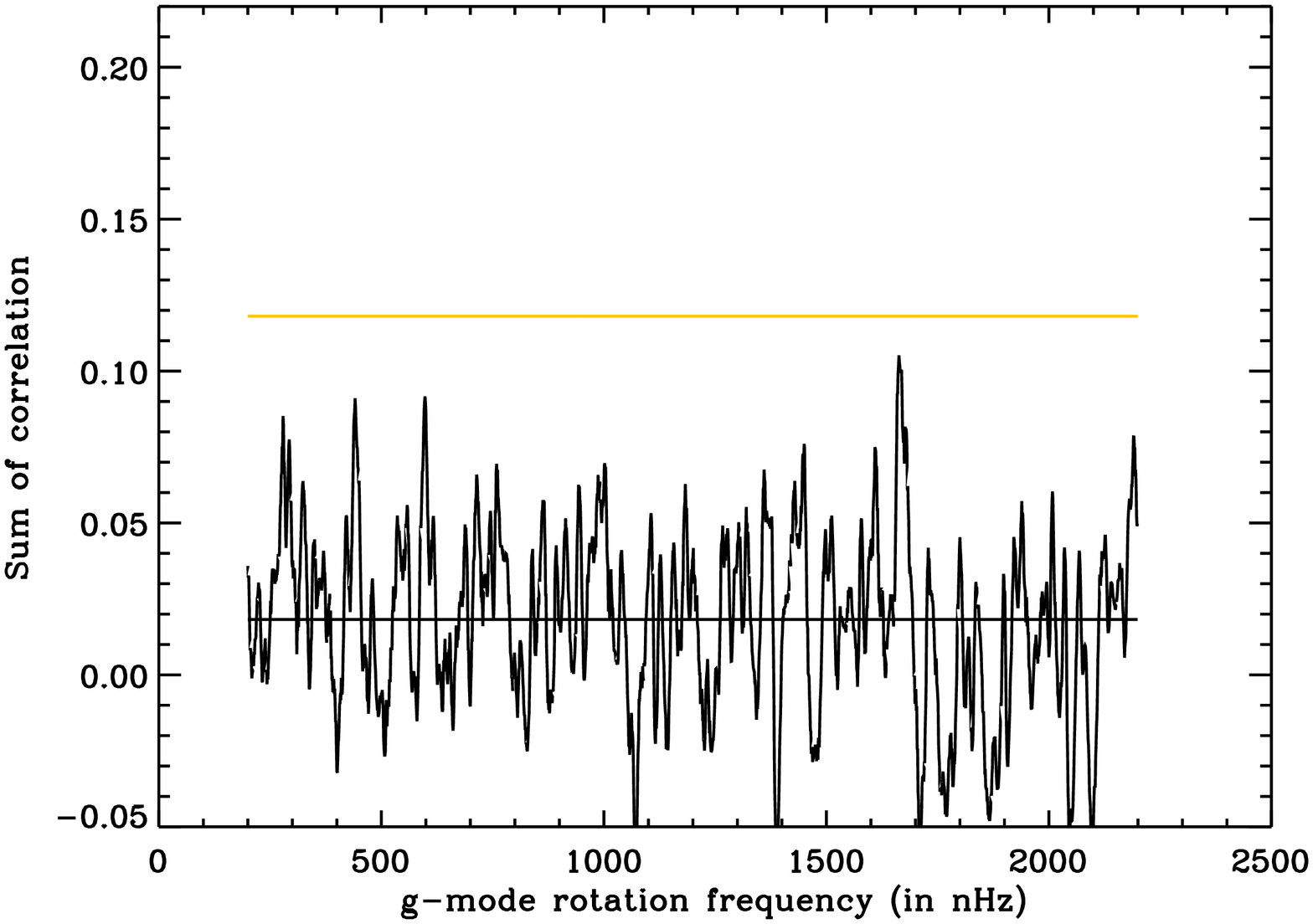}}
\hbox{
\includegraphics[width=6 cm,angle=90]{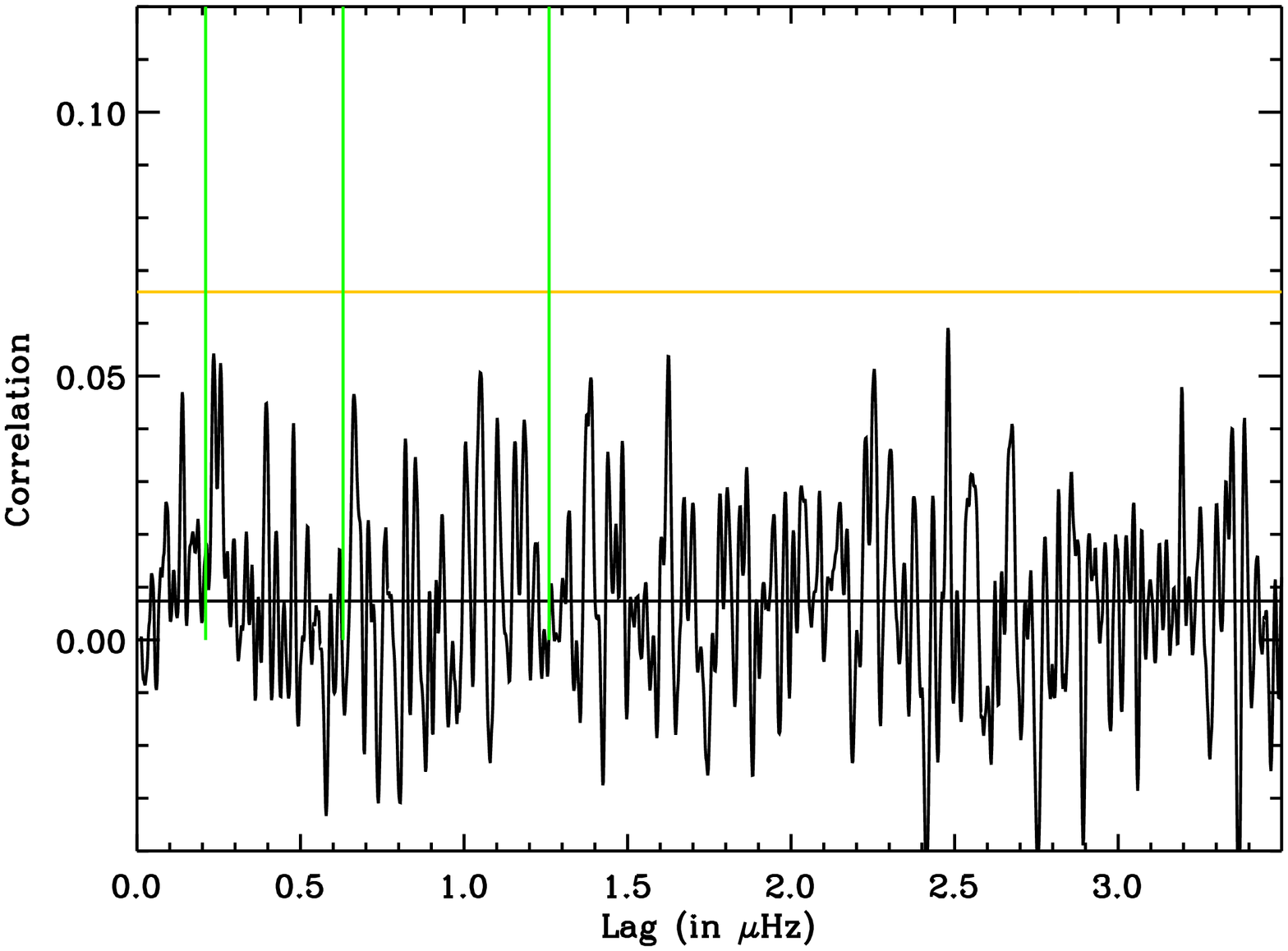}
\includegraphics[width=6 cm,angle=90]{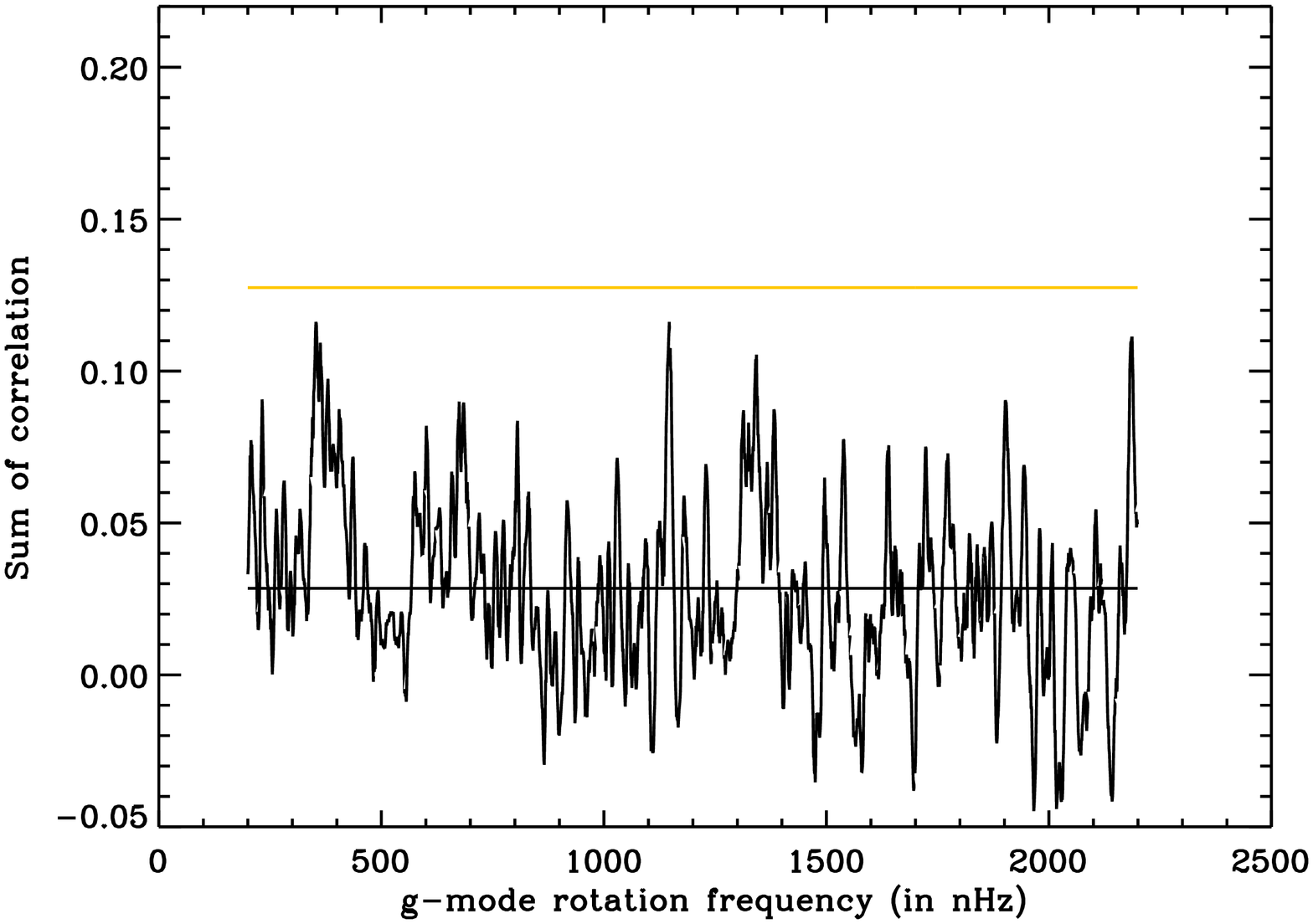}}
}}
\caption{(left) Autocorrelation of the power spectrum as obtained by \citet{Fossat2017} as a function of frequency lag for the two different GOLF photomultipliers sampled at 20 s for PM1 (top) and PM2 (bottom).  The green vertical lines correspond to frequencies at 210 nHz, 630 nHz and 1260 nHz.  The black line indicates the mean value, while the orange line indicates the 10\% probability level that there is at least one peak due to noise in the window; the level is 3.41 $\sigma$.  (right) Sum of the correlation for $l=1$, and $l=2$ modes as obtained by \citet{Fossat2017} as a function of rotation frequency for the two different photomultipliers of GOLF sampled at 20 s for PM1 (top) and PM2 (bottom).  The black line indicates the mean value, while the orange line indicates the 10\% probability level that there is at least one peak due to noise in the window; the level is 3.43 $\sigma$.}
\label{pm}
\end{figure*}

The original claim in F17 was made using the average of the two GOLF photomultipliers.  Neither F17 nor \citet{Schunker2018} studied the impact of the different photomultipliers upon the g-mode detection claim.  Here we used the newly calibrated GOLF data of each photomultiplier with the same observation duration as in Paper~I.  Figure~\ref{pm} shows the results for the two different photomultipliers.  There is no peak above the 10\% threshold in any panel.  Since this work was not done in F17, we try to forecast the signal-to-noise ratio in a single photomultiplier from a time series two times shorter instead.  F17 gave a signal-to-noise ratio for an observation time of 8.25 years that is typically about 2.6 $\sigma$ either for the peak at 210 nHz, or the peak at 1280 nHz.  From Fig.~\ref{pm}, it is clear that these peaks are nowhere near this signal-to-noise ratio.  Here we conclude that the g-mode detection claim of F17 is not confirmed with either individual GOLF photomultiplier.  Although we show only the results for the 60-s sampling, we confirm that there is no detection either for the 20-s and 80-s sampling.


\subsection{Different sub-series}
\begin{figure*}[!]
\centerline{
\vbox{
\hbox{
\includegraphics[width=6 cm,angle=90]{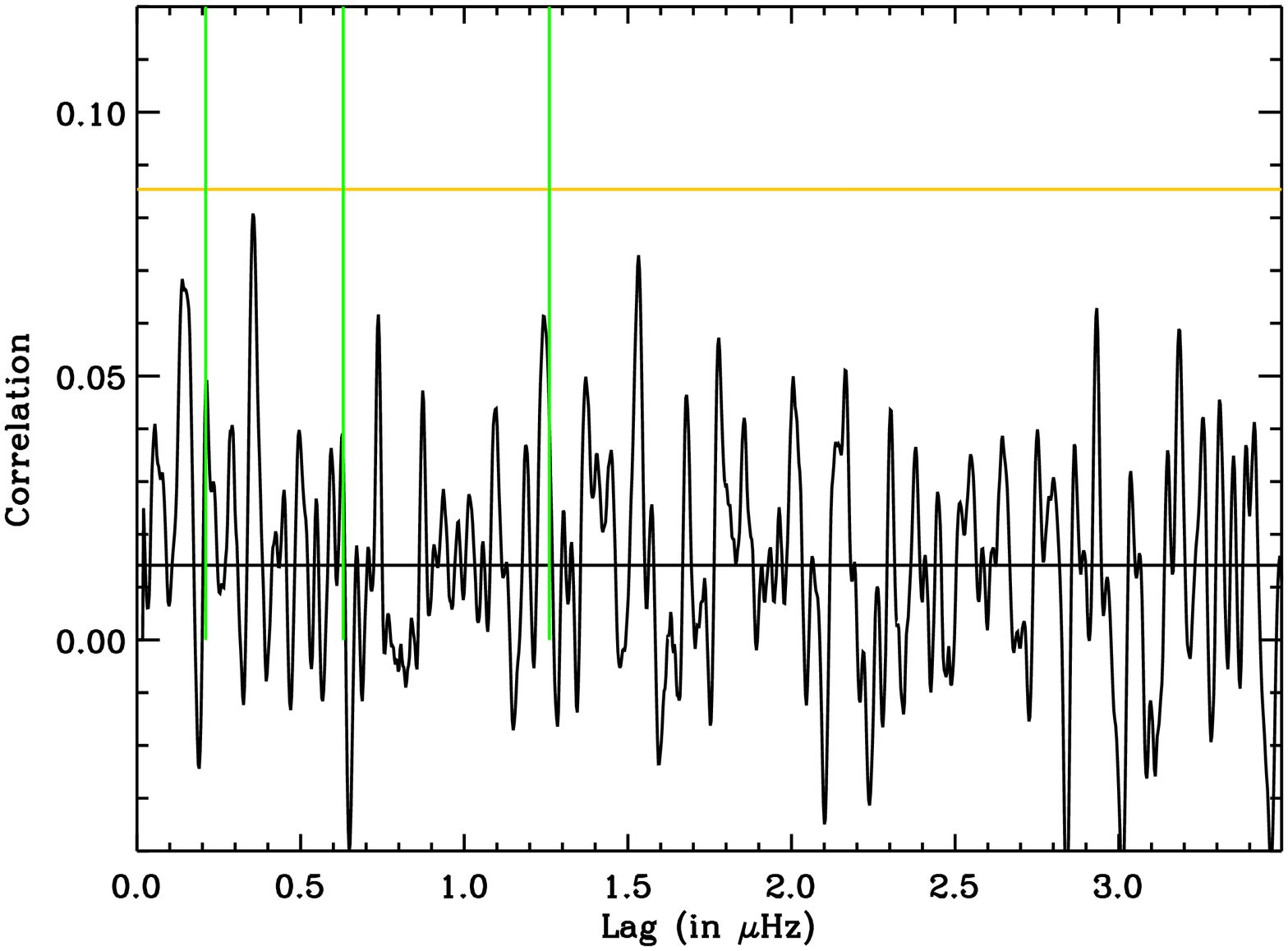}
\includegraphics[width=6 cm,angle=90]{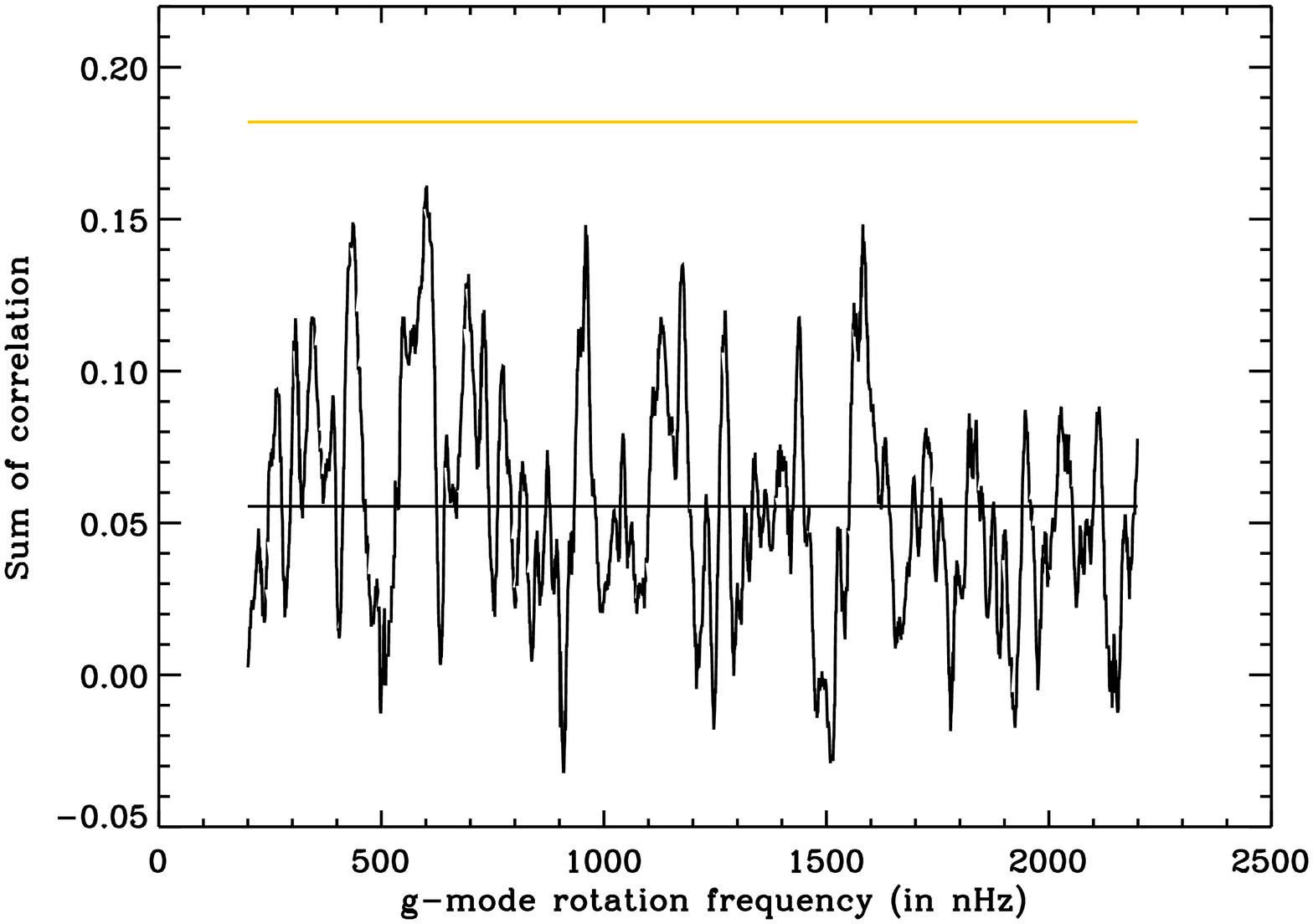}}
\hbox{
\includegraphics[width=6 cm,angle=90]{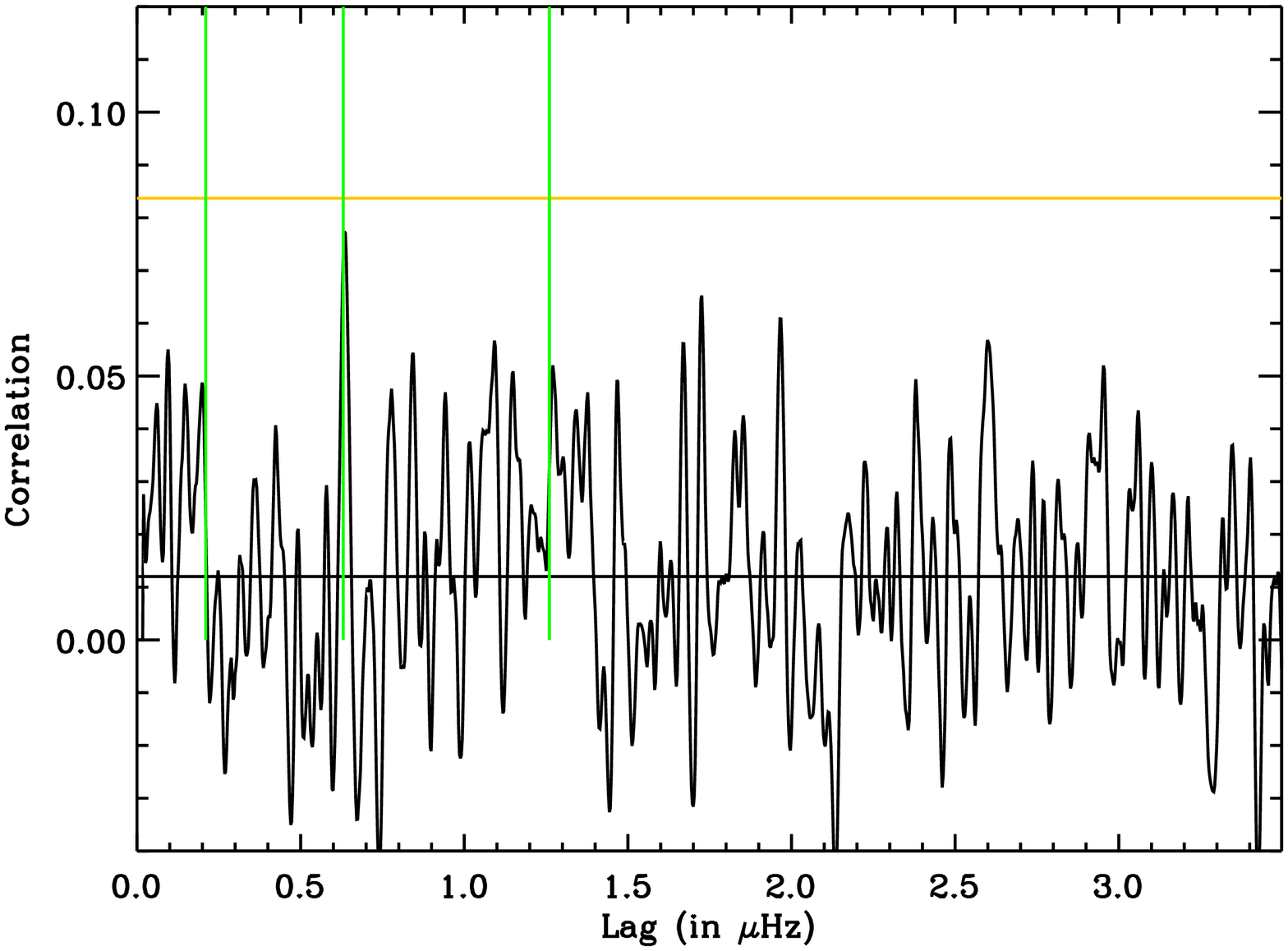}
\includegraphics[width=6 cm,angle=90]{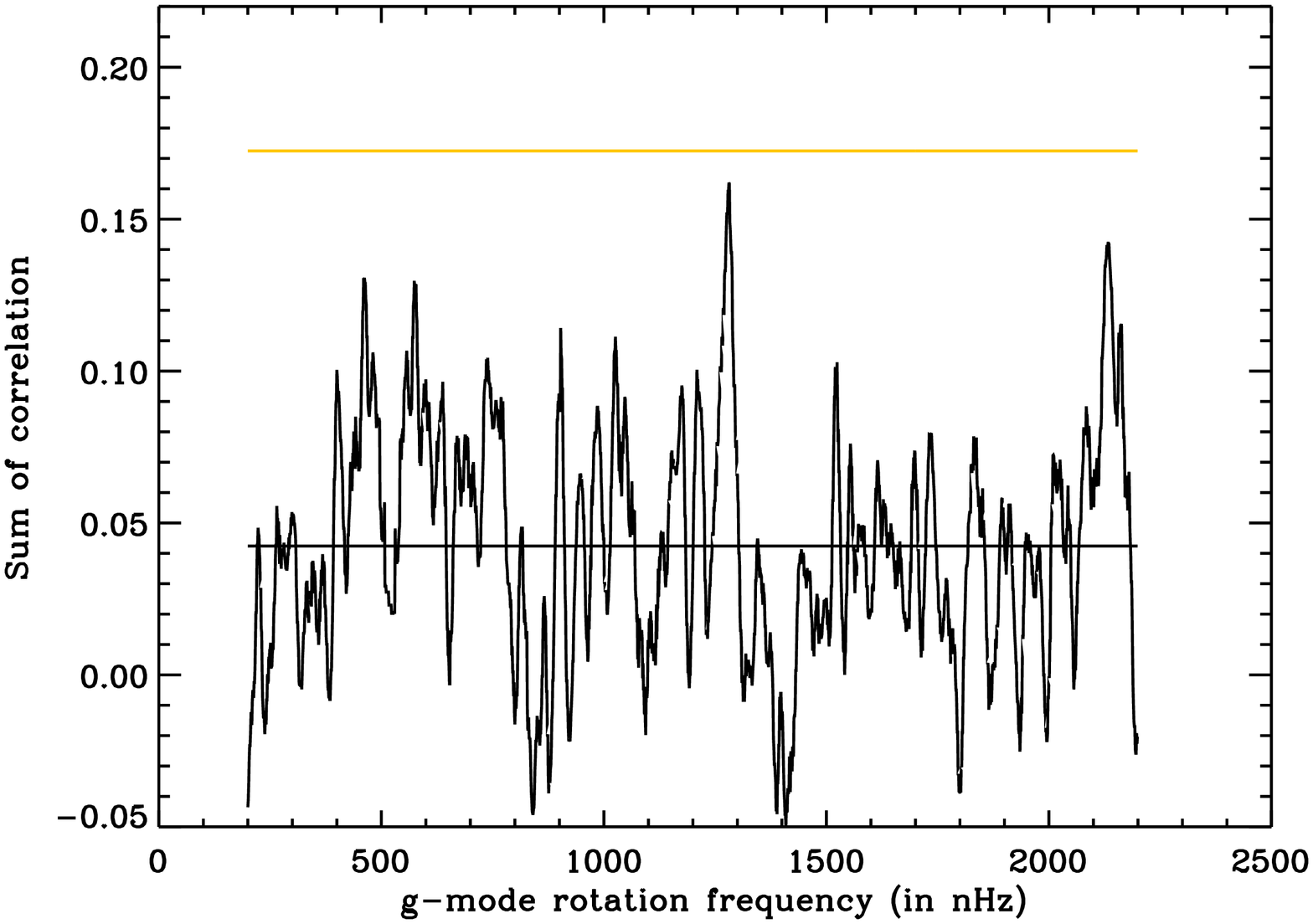}}
\hbox{
\includegraphics[width=6 cm,angle=90]{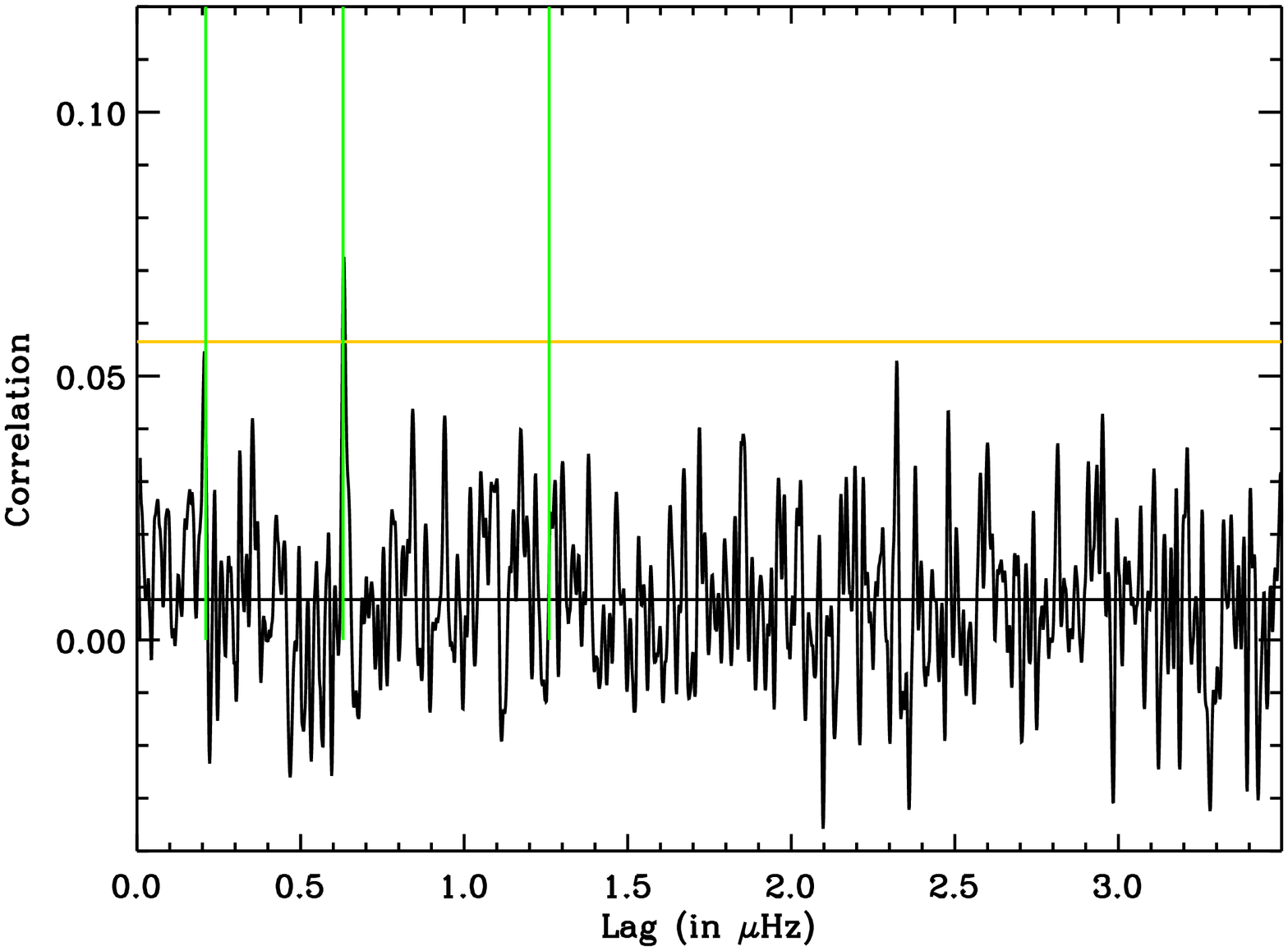}
\includegraphics[width=6 cm,angle=90]{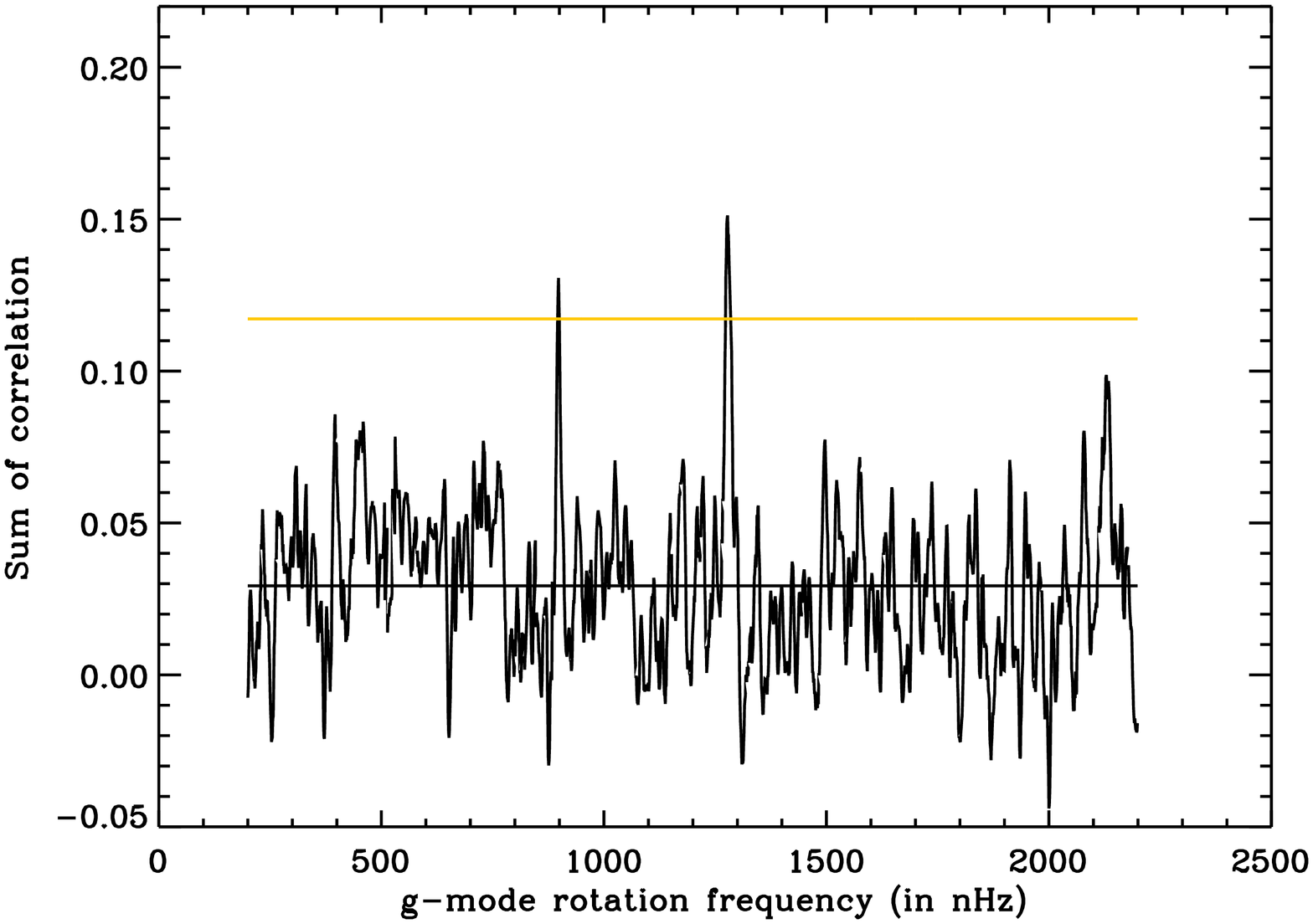}}
}}
\caption{(left) Correlation of the power spectrum as obtained by \citet{Fossat2017} as a function of frequency lag for three different time series of the average of photomultipliers PM1 and PM2 of GOLF sampled at 80 s for an observation time of 11 years starting April 1st 1996 (top), for an observation of  11 years starting April 1st 2007 (middle) and an observation time of 22 years (bottom).  The green vertical lines correspond to frequencies at 210 nHz, 630 nHz and 1260 nHz.  The black line indicates the mean value, while the orange line indicates the 10\% probability level that there is at least one peak due to noise in the window; the level is 3.29 $\sigma$ for 11 years, and 3.48 $\sigma$ for 22 years.  (right) Sum of the correlation for $l=1$, and $l=2$ modes as obtained by \citet{Fossat2017} as a function of rotation frequency  three different time series of the average of photomultipliers PM1 and PM2 of GOLF sampled at 80 s for an observation time of 11 years starting April 1st 1996 (top), for an observation of  11 years starting April 1st 2007 (middle) and an observation time of 22 years (bottom).  The black line indicates the mean value, while the orange line indicates the 10\% probability level that there is at least one peak due to noise in the window; the level is 3.4 $\sigma$}
\label{subseries}
\end{figure*}

The original claim in F17 was made using an observation time of 16.5 years.  F17 reported that the signal-to-noise ratio of the 210-nHz peak and of the 1280-nHz peak increase with the observation time.  With our new calibration (Paper I) we could extend the observation time to 22 years, and to two independent time series of 11 years.  Figure~\ref{subseries} shows the results for the three different time series.  With a time series of 11 years, F17 reported a signal-to-noise ratio of 3.5 $\sigma$ for the 210-nHz and 1280 nHz peak.  In comparison, in Fig.~\ref{subseries} the signal-to-noise ratio is twice smaller for the 210-nHz peak, but still increases with the longer time series even though it remains below the 10\% line.  On average the signal-to-noise ratio is also twice smaller for the 1280-nHz peak but also increases with the longer times series to be above the 10\% line.  On the other hand, another peak shows up at 900 nHz in the sum of the correlation that was not mentioned in F17.  The signal-to-noise ratio of the 900-nHz peak is also higher for the longer times series.  We conclude that the signal-to-noise ratio of the peaks mentioned in F17 increases with the observation time but that we cannot confirm detection in the 11-year long sub-series, and that only the peak at 630 nHz is detected in the longest times series.


\subsection{Different instruments}
\begin{figure*}[!]
\centerline{
\vbox{
\hbox{
\includegraphics[width=6 cm,angle=90]{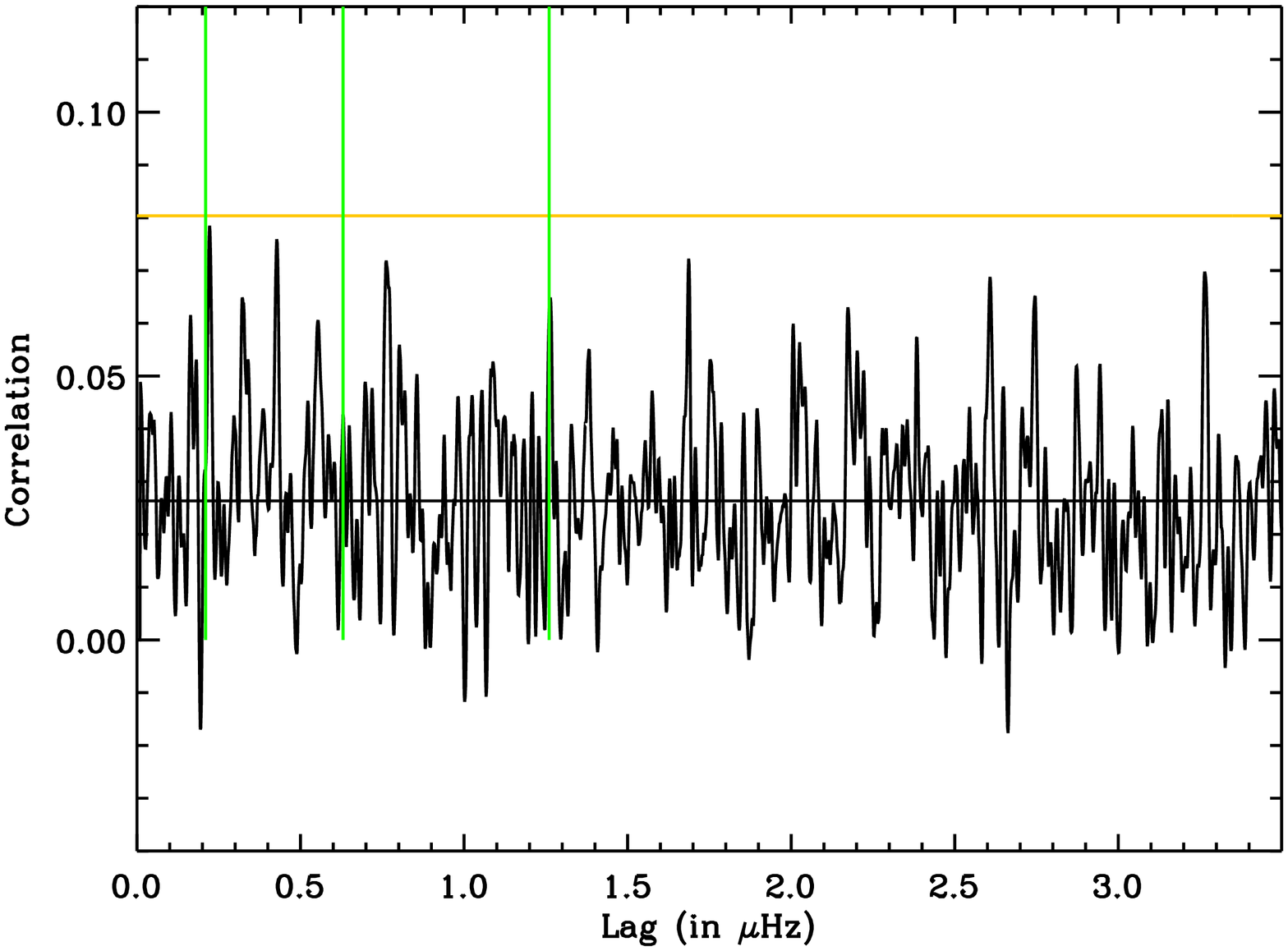}
\includegraphics[width=6 cm,angle=90]{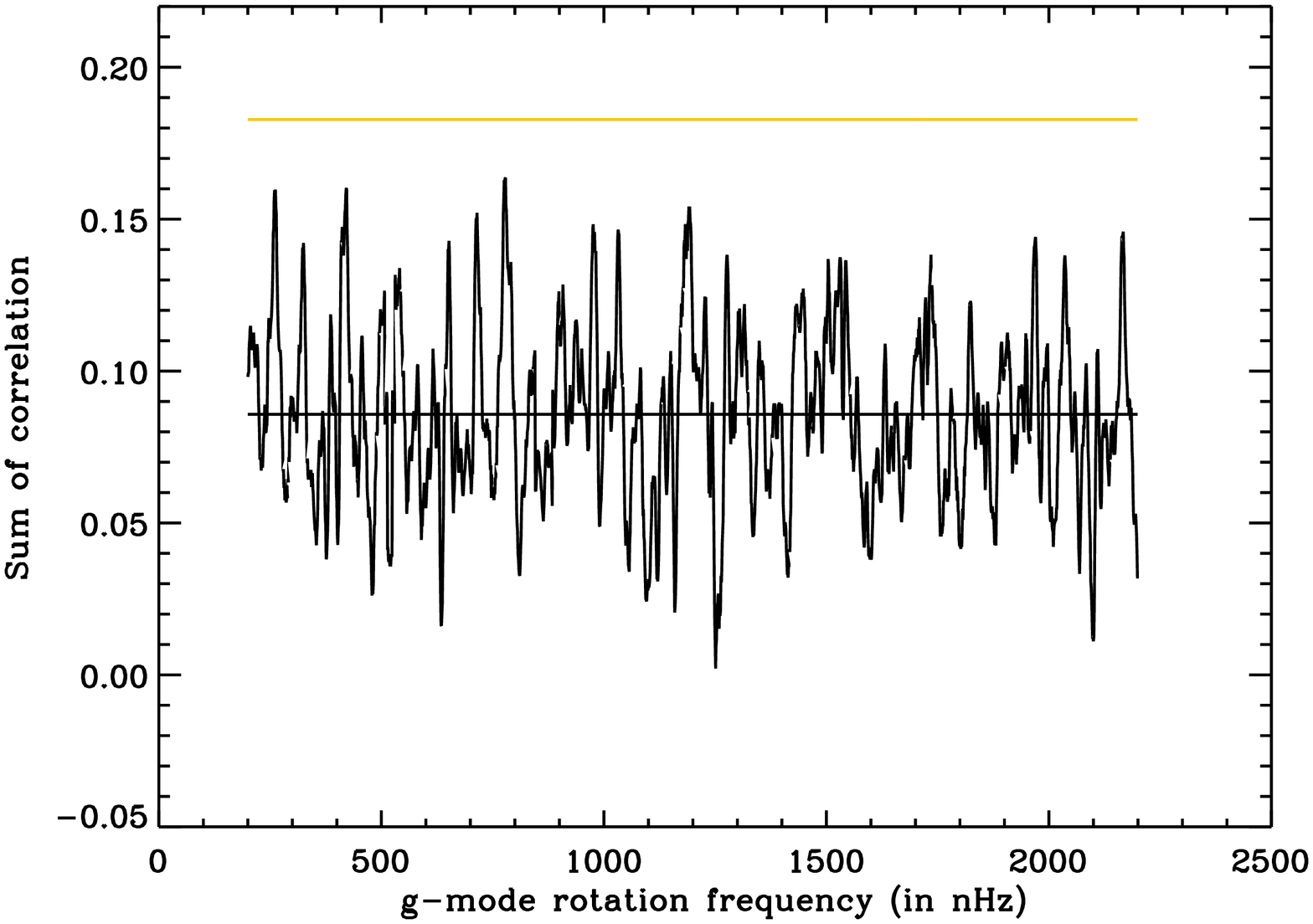}}
\hbox{
\includegraphics[width=6 cm,angle=90]{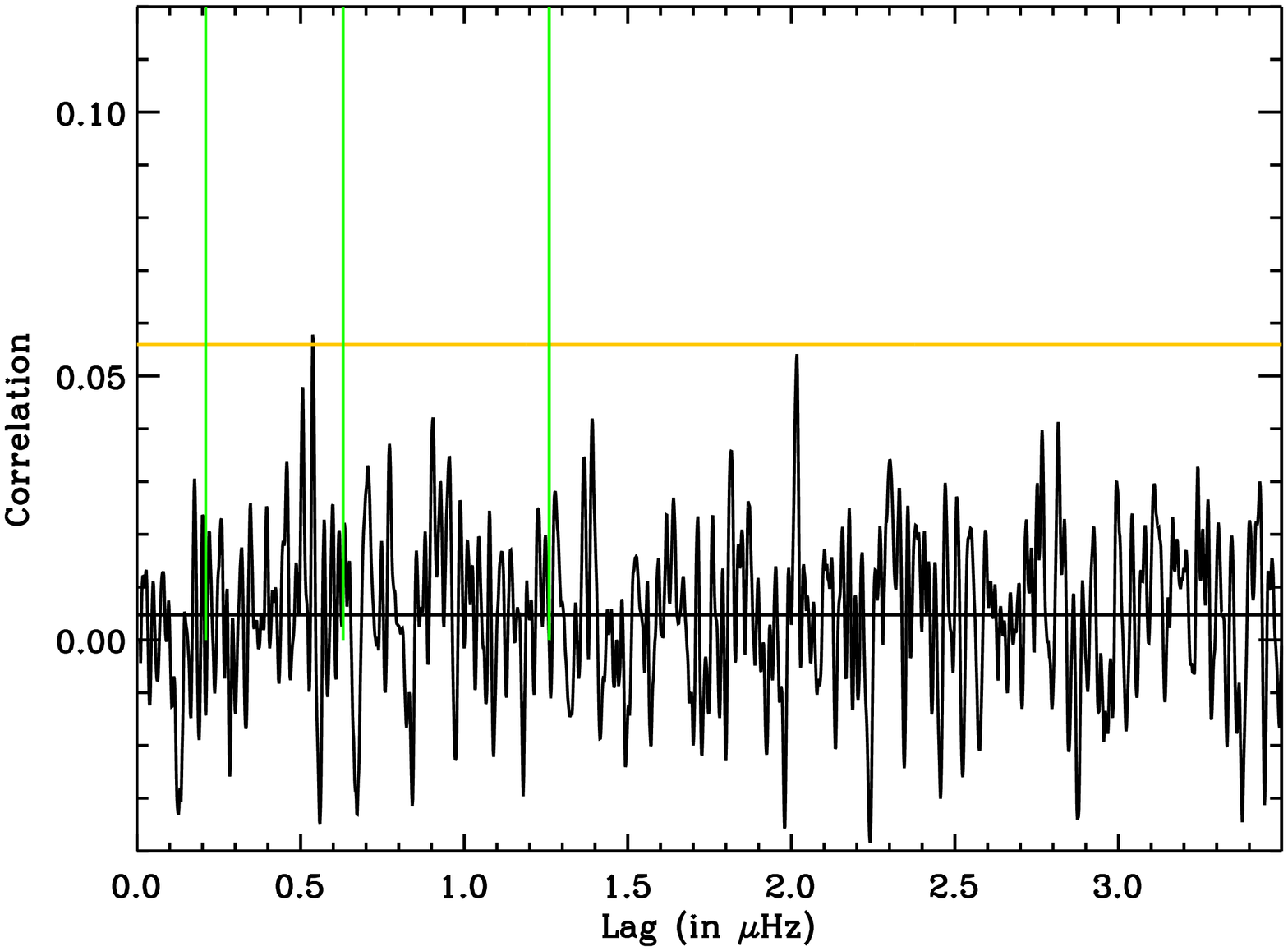}
\includegraphics[width=6 cm,angle=90]{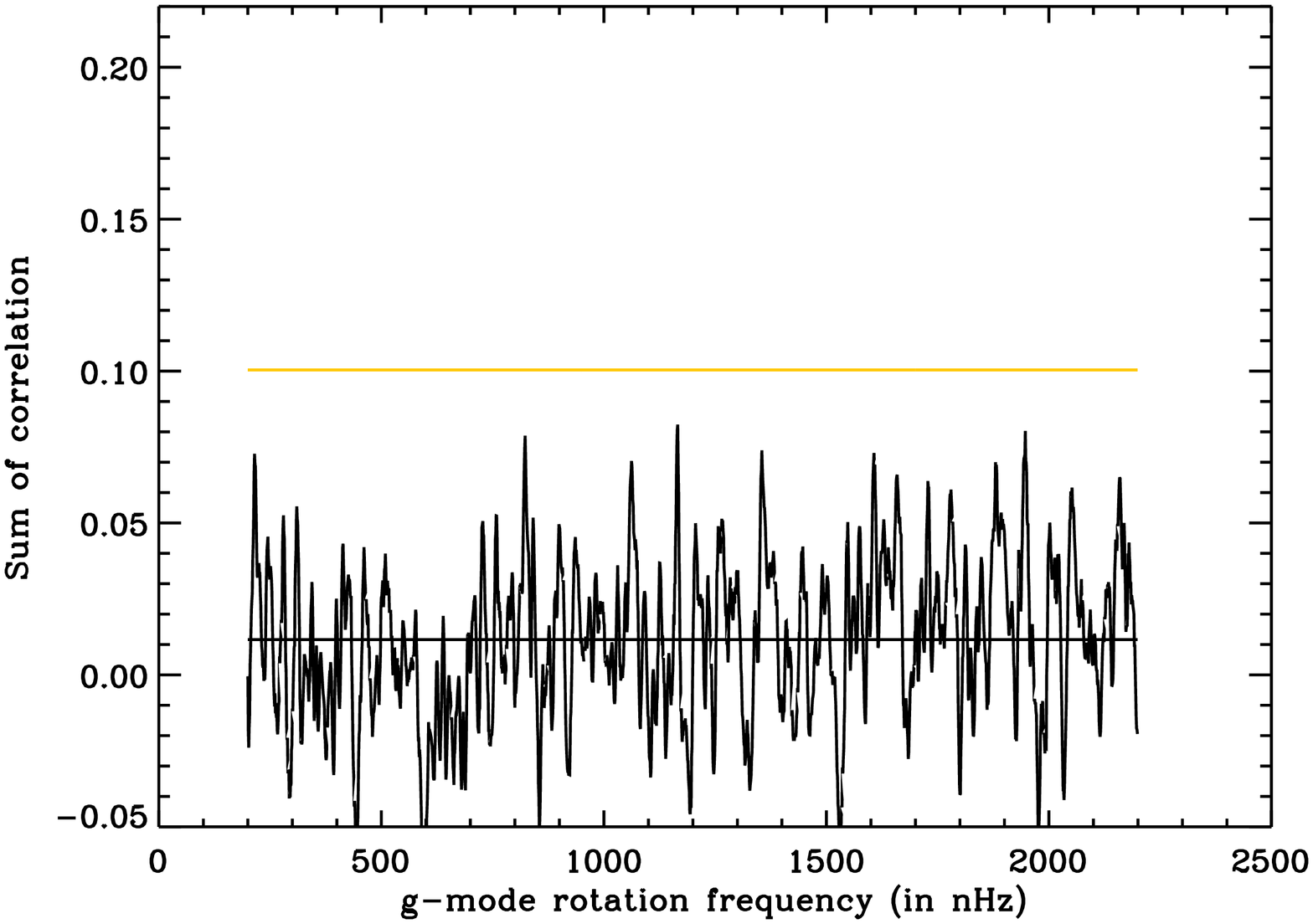}}
}}
\caption{(left) Autocorrelation of the power spectrum as obtained by \citet{Fossat2017} as a function of frequency lag for the GONG instrument (top) and the BiSON instrument (bottom).  The green vertical lines correspond to frequencies at 210 nHz, 630 nHz and 1260 nHz.  The black line indicates the mean value, while the orange line indicates the 10\% probability level that there is at least one peak due to noise in the window; the level is 3.46 $\sigma$ for GONG,  and 3.48 $\sigma$ for BiSON.  (right) Sum of the correlation for $l=1$, and $l=2$ modes as obtained by \citet{Fossat2017} as a function of rotation frequency for the GONG instrument (top) and the BiSON instrument (bottom).  The black line indicates the mean value, while the orange line indicates the 10\% probability level that there is at least one peak due to noise in the window; the level is 3.43 $\sigma$.}
\label{instru}
\end{figure*}

We used data from other instruments such as the Global Oscillation Network Group\footnote{GONG, See \citet{Harvey1996}} and the Birmingham Solar Oscillation Network\footnote{BiSON, See \citet{WJC1996}}.  For GONG we used the full-disk integrated velocity with a start date of May 1st, 1996 lasting 21.3 years.  For BiSON, we used the performance-check data \footnote{Data available on bison.ph.bham.ac.uk/} truncated to starting on May 1st, 1996 lasting 20 years.  Figure~\ref{instru} gives the result for the GONG and BiSON instruments.  There is no confirmation of the g-mode detection of F17 at a 10\% level (>38\% for the posterior probability of H$_0$).


\subsection{Summary}
In addition to the work presented here, we also checked the results found by \citet{Schunker2018} for the change of the start time.  We also found that the detection cannot be confirmed when shifting the start date by 2 hours, and that there is a pseudo-periodicity of 4 hours due to the construction of the time series having a 4-hour sampling.  We also used a different fitting strategy of the p-mode peak at 4.1 hours to derive that the {\rm best} fit minimising the rms scatter of the round-trip travel time was indeed a 2nd-order polynomial.  Here we give a short summary of the impact of various parameter on the g-mode detection claim of F17:
\begin{itemize}
\item {\bf Shifted starting time}: as also found by \citet{Schunker2018}, a shift of 2 hours suppresses the g-mode detection. 
\item {\bf Cadence}: \citet{Schunker2018} reported that a cadence different than 4 h (3 h to 5 h) also suppresses the g-mode detection.
\item {\bf Different fit function}: as also found by \citet{Schunker2018}, other function, such as a Gaussian would also impact the g-mode detection
\item {\bf Power spectrum smoothing}: \citet{Schunker2018} reported that the smoothing window also impacts the g-mode detection
\item {\bf Sampling}: might affect the detection as the heights of the 3 peaks in the correlation (210 nHz, 630 nHz, 1260 nHz) depend upon the sampling.  The peak at 1280 nHz in the sum of the correlation is less dependent upon the sampling
\item {\bf Photomultipliers}: no confirmed detection from either photomultiplier
\item {\bf Sub-series}: detection confirmed for the same 16.5 years time series as F17.  None of the two 11-y long time series confirms the detection, while the 22-y long time series confirmed detection of some peaks but not of the main 210-nHz peak.  An additional peak shows up in the sum of the correlation at 900 nHz
\item {\bf Instruments}: no confirmed detection in other instruments (GONG, BiSON) measuring solar radial velocities.
\end{itemize}
Although the detection claimed in F17 is not always reproducible, we must point out that it is quite remarkable to have obtained 2 peaks above the 10\% threshold in Fig.~\ref{sampling}, which was not obtained in F17.  It clearly shows that the levels of these 2 peaks are very sensitive to the details of the procedure.

\section{Reproducing the analysis of FS18}
\subsection{g-mode asymptotic model}
In FS18, the analysis done in F17 was pushed further by correlating an asymptotic model of g-mode periods with the power spectrum of the time series of round-trip travel time.  The asymptotic model of g-mode period $P_{n,l}$ can be found in \citet{JPGB1986} as:
\begin{equation}
P_{n,l}=\frac{P_0}{\sqrt{l(l+1)}}\left(n+\frac{l}{2}-\frac{1}{4}-\theta\right)+\frac{P_0^{2}}{P_{n,l}}\frac{l(l+1)V_1+V_2}{l(l+1)}
\end{equation}
where $n$ is the order of the mode, $l$ is the degree of the mode, and $P_0$, $V_1$, $V_2$ and $\theta$ are all related to the solar model.  Equation (2) can be solved for $P_{n,l}$
and can then be rewritten as:
\begin{equation}
P_{n,l}=\frac{P_0}{\sqrt{l(l+1)}}\left[\left(n+\frac{l}{2}-\frac{1}{4}-\theta\right)+\frac{(l(l+1)V_1+V_2)}{(n+\frac{l}{2}-\frac{1}{4}-\theta)} \right]+{\rm O}\left(\frac{1}{n^3}\right)
\end{equation}
where ${\rm O}\left(\frac{1}{n^3}\right)$ is a quantity of order $\frac{1}{n^3}$.  Assuming that $n \gg 1$, we can then rewrite $P_{n,l}$ as:
\begin{equation}
P_{n,l}=P_{n_0,l}+P_l\left(n-n_0\right)+\alpha\left(\frac{1}{n}-\frac{1}{n_0} \right)+{\rm O}\left(\frac{1}{n^3}\right)
\end{equation}
where $P_l=\frac{P_0}{\sqrt{l(l+1)}}$ is the asymptotic periodicity of the modes of degree $l$, $n_0$ is an arbitrary
value of g-mode order and $\alpha$ is given by:
\begin{equation}
\alpha=P_l (l(l+1)V_1+V_2)
\end{equation}
Finally, we can rewrite Eq. (3) as:
\begin{equation}
P_{n,l} \approx P_{\rm min,l}+P_l\left(n-n_0\right)+\frac{\alpha}{n}
\end{equation}
This equation is the same as Eq. (14) in F17 except that $\alpha$ in our formula is directly related to $P_l$ while in F17 it is taken as a free parameter $\alpha_l$.  In F17, the model of the g-mode frequency is then given by:
\begin{equation}
\nu_{n,l,m}=1/P_{n,l}+m \nu_{l}
\end{equation}
where $\nu_{l}$ is the rotational mode splitting for modes of degree $l$ assumed to be independent of $n$ and $m$.  In addition, each mode of tesseral order $m \ne 0$ of a given $l$ is assumed to be observed with a weight $w_{l,m}$ less than 1. In summary, for $l=1$ there are five independent parameters defining the model: $P_1$, $P_{\rm min,1}$, $\alpha_1$, $\nu_1$ and $w_{1,1}$; while for $l$=2, there are six independent parameters defining the model: $P_2$, $P_{\rm min,2}$, $\alpha_2$, $\nu_2$, $w_{2,1}$ and $w_{2,2}$.  Here we note that F17 assumed that the parameters so defined are independent of each other, while as shown by Eq. (5) the $P_l$ and $\alpha$ depend upon $P_0$.  Here in order to reproduce, the findings of FS18 we followed the prescription of F17 for the independence of the parameters.

\subsection{Optimisation of the correlation}
The model given by Eqs. (6) and (7) is then correlated with the smoothed power spectrum of time series of round-trip travel time.  Here we note that the power spectrum is smoothed over 2 bins in FS18 rather than the 6 bins of F17.  The computed correlation is made symmetrical by adding a flipped version of the correlation to itself.  Then in FS18, the correlation value at 0 and at the splitting lag are optimised or fine tuned by adjusting all 5 parameters for $l=1$ and all 6 parameters for $l=2$.  The optimisation procedure in FS18 is not detailed but done by hand (Fossat, private communication, 2018).  In the course of the investigation, it appeared that the finding of a maximum could not be made possible by any simple optimisation routine mainly because the shape of the correlation as a function of the free parameters was not following a simple paraboloid shape.  In order, to have the highest correlation possible, we decided to automate the optimisation procedure.  The brute force procedure that we used for the optimisation is as follows:
\begin{enumerate}
\item Shoot at random the 5 (or 6) parameters within a given hypersphere
\item Compute the symmetrical correlation of the model spectrum with the smoothed power spectrum
\item Compute the sum of the correlation at lag=0 and at lag=$\nu_l$
\item Keep shooting until the sum does not increase by more than 0.5\% (make sure there are at least 10$^5$ shots).  Reduce hypersphere by 10\%, each time a higher sum is found
\item Repeat at most 5 millions times Step 1 to 4 (unless the condition in Step 4 is obtained)
\end{enumerate}

%

\subsection{Noise}
\begin{figure*}[!]
\centerline{
\vbox{
\hbox{
\includegraphics[width=6 cm,angle=90]{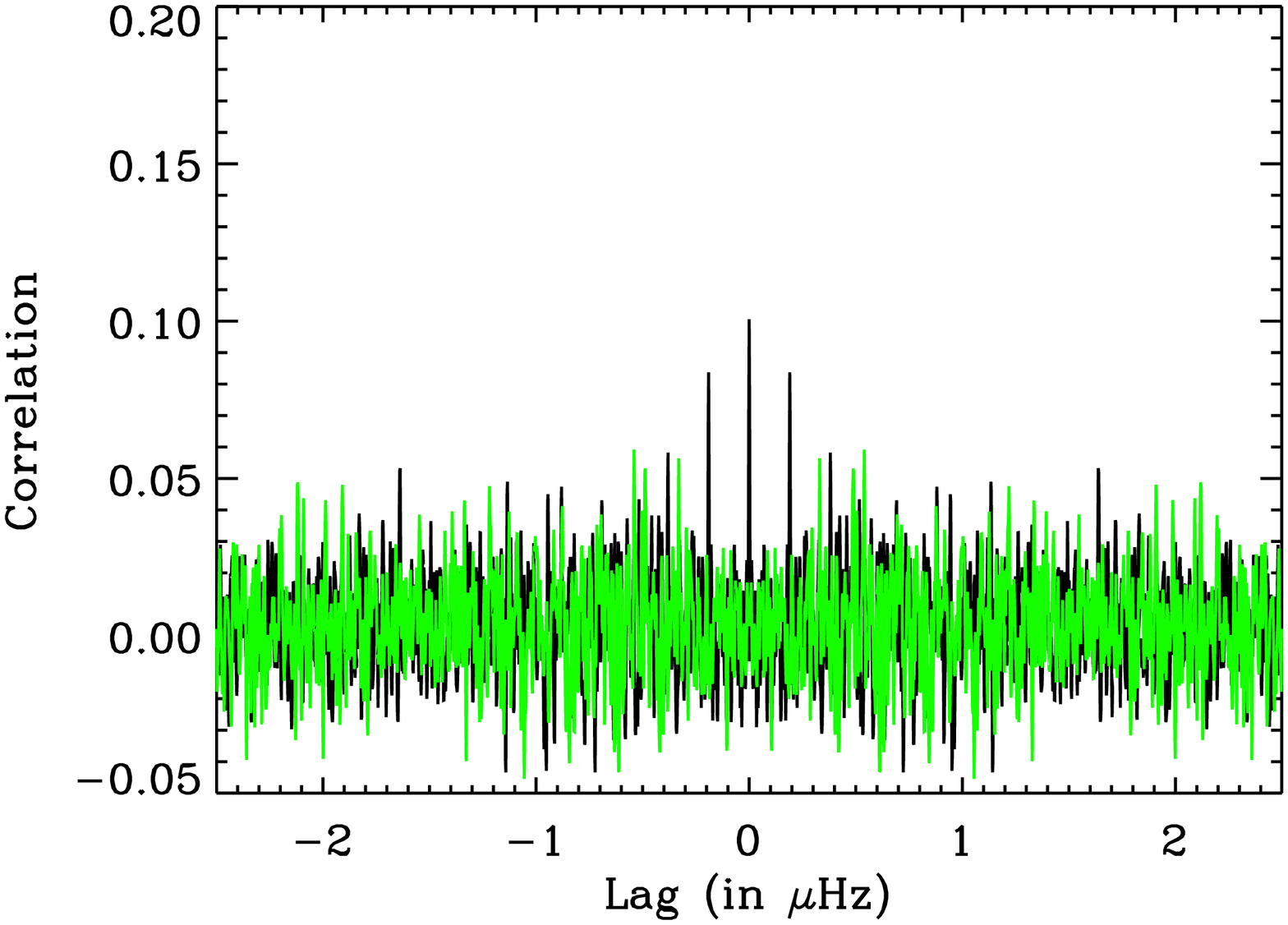}
\includegraphics[width=6 cm,angle=90]{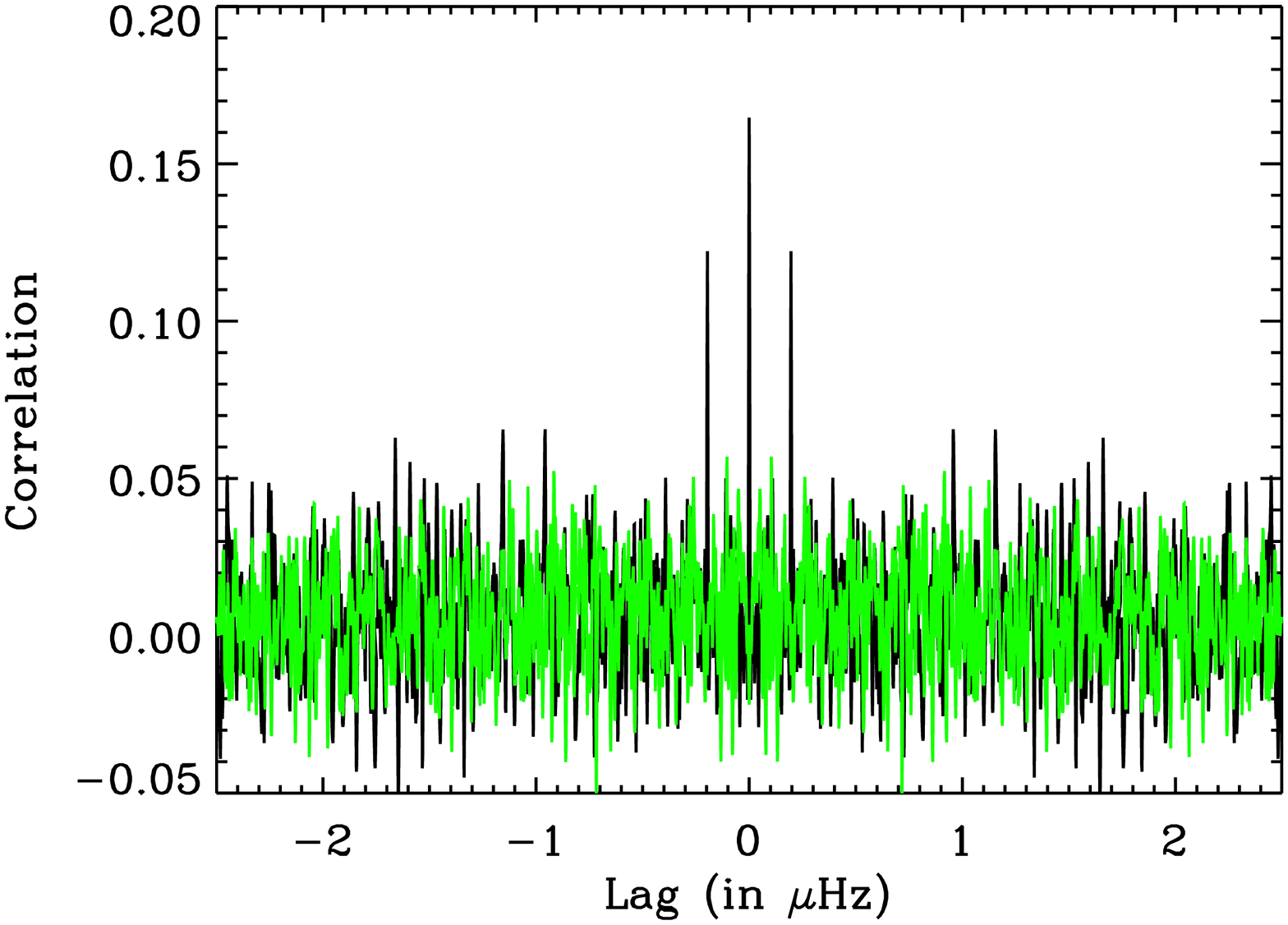}}
\hbox{
\includegraphics[width=6 cm,angle=90]{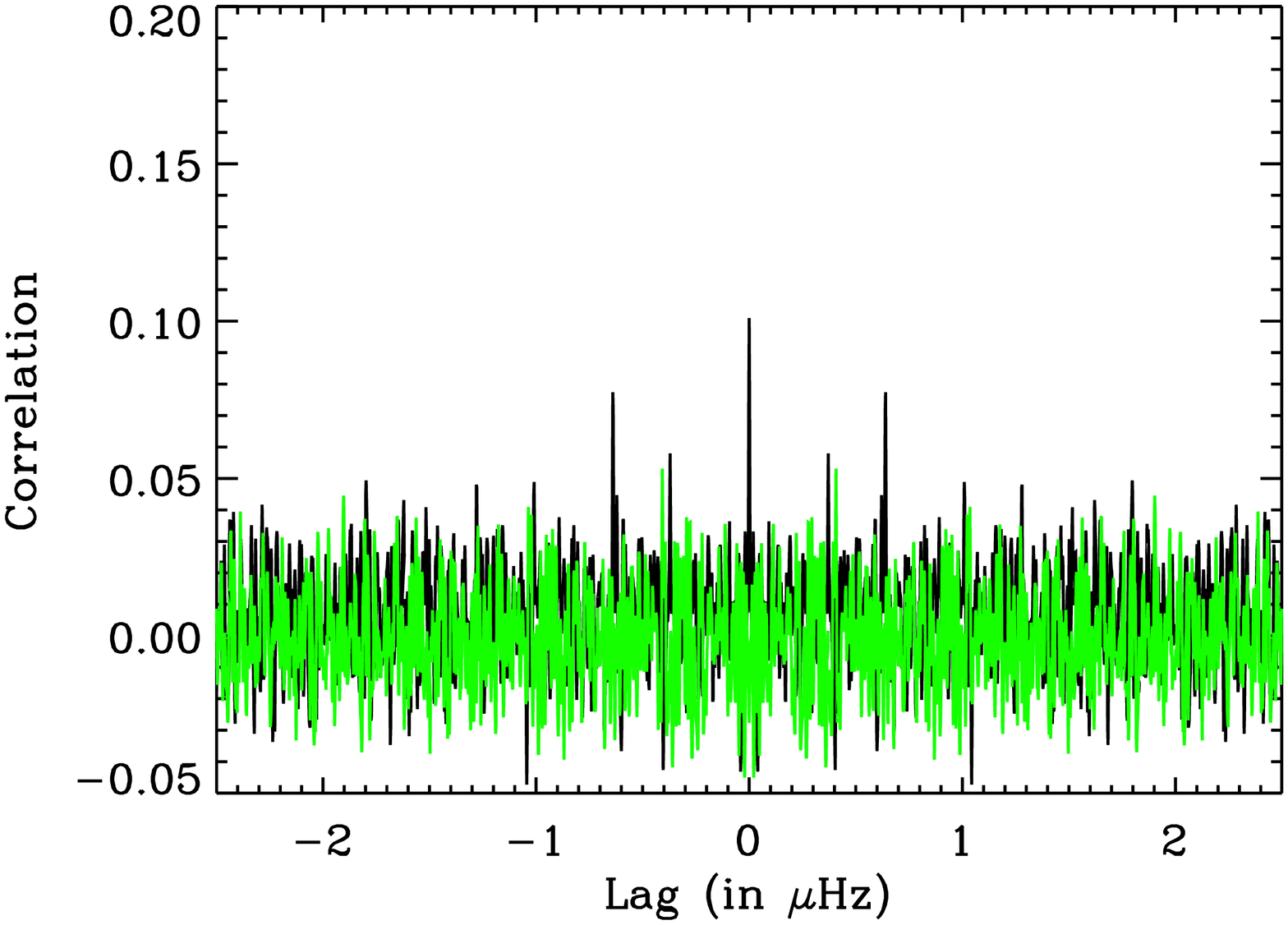}
\includegraphics[width=6 cm,angle=90]{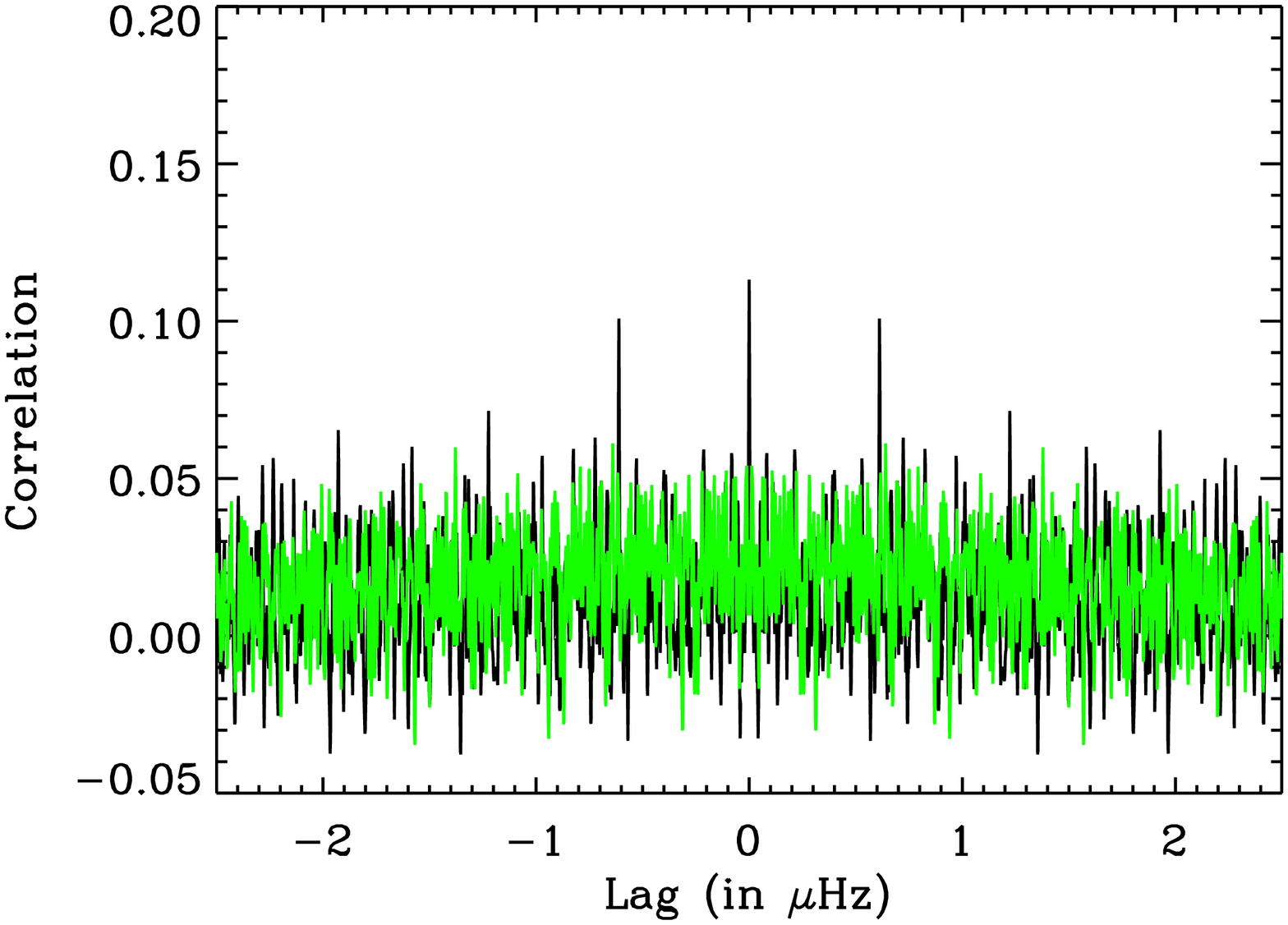}}
\hbox{
\includegraphics[width=6 cm,angle=90]{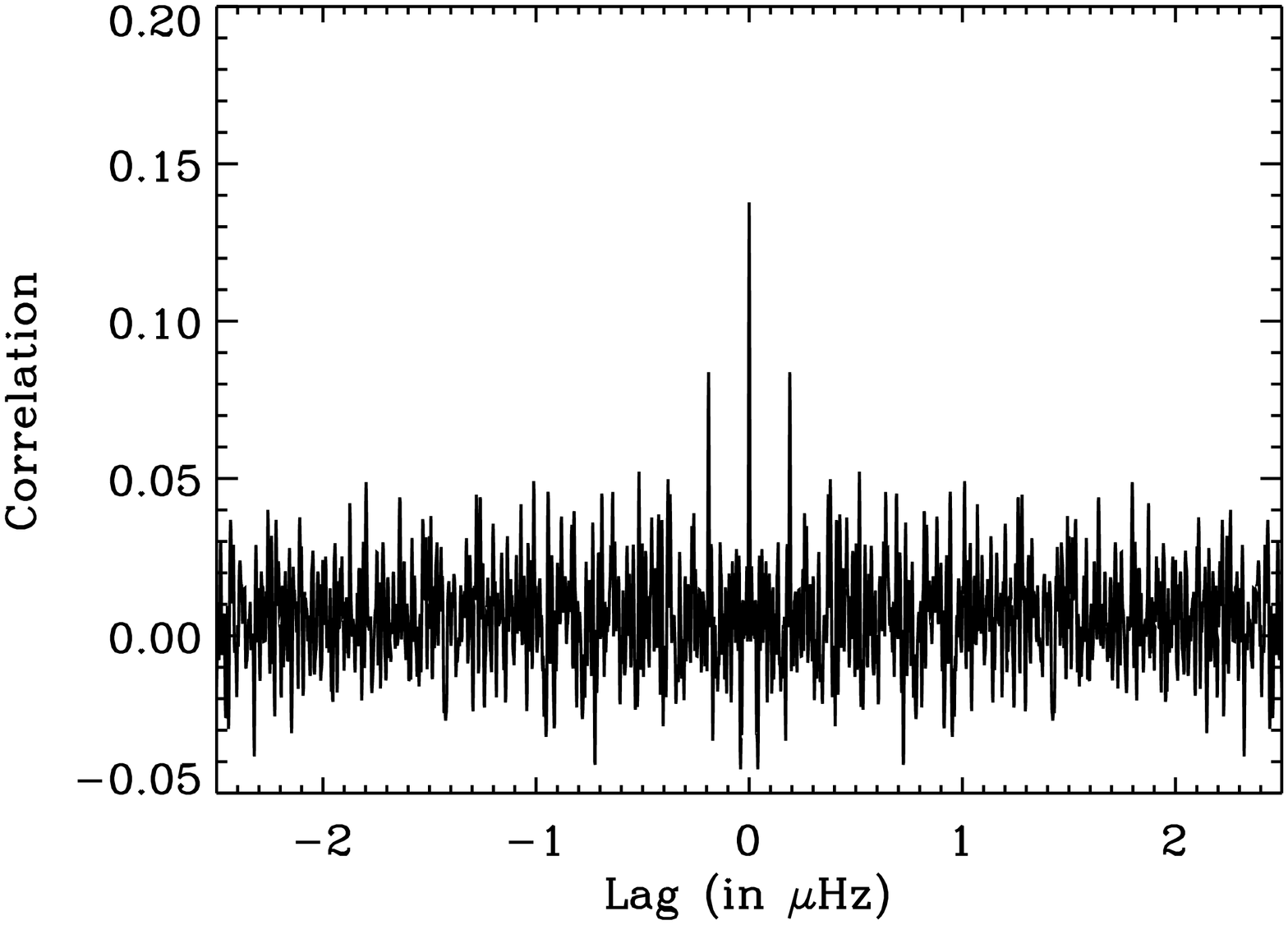}
\includegraphics[width=6 cm,angle=90]{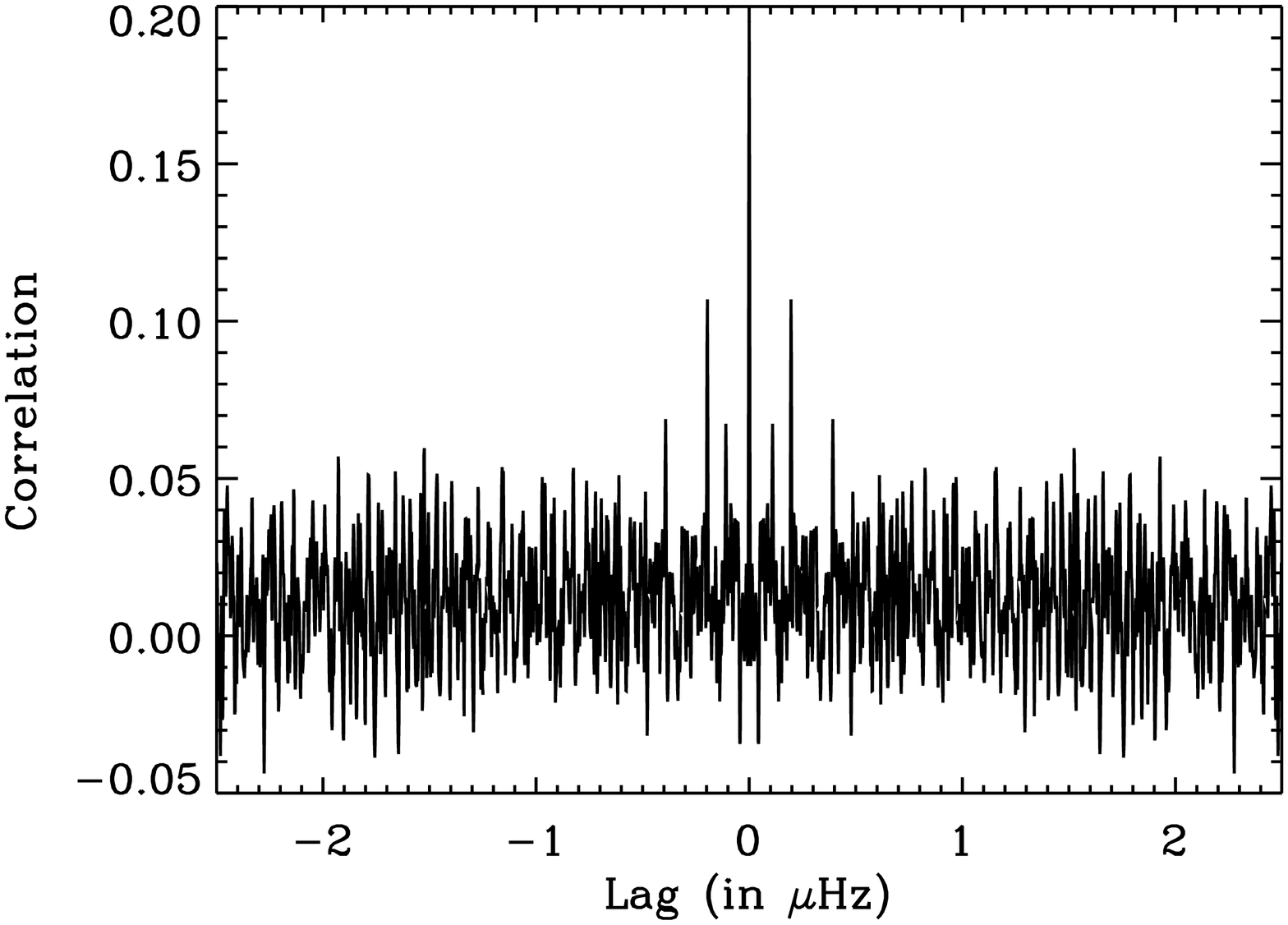}}
}}
\caption{Correlation of the g-mode model spectrum with the power spectrum of a simulated time series of round-trip travel time: (Top) For $l=1$ only (similar to Fig. 1 in FS18).  The black line is for the result after optimisation while the green it is before optimisation ; (Middle) For $l=2$ only.  The black line is for the result after optimisation while the green it is before optimisation; (Bottom) For $l=1$ and $l=2$ (similar to Fig. 2 in FS18).  (left) For the median value of the sum of the correlations at lag=0 and lag=$\nu_1$ obtained over the 2000+ simulations.  (right) For the maximum value of the sum of the correlations at lag=0 and lag=$\nu_1$ obtained over the 2000+ simulations.
}
\label{correls}
\end{figure*}

In order to understand the correlation results, we applied the procedure described above to pure noise.  We generated 36132 data points sampled at 4 h for simulating the time series of 16.5 years of data of F17.  Under the null hypothesis, we assumed a white noise with Gaussian statistics with a mean of 0 and rms of 52 s from which we deduced after Fourier transform the power spectrum.  For the simulation, we then generated a bit more than 2000 power spectra to which we applied the optimisation procedure described above.  For the search, the parameters used for the optimisation were generated in a hypersphere given in Table~\ref{noise_1} and Table~\ref{noise_2}.  The optimisation of the correlation with the model spectrum is done only for $l=1$ for comparison with Fig. 1 in FS18.  Then we optimise for $l=2$, and add half of the model spectrum of $l=2$ to the model spectrum of $l=1$ for getting the correlation shown in Fig. 2 in FS18 (the weighting used in F17 is 0.43 instead of half). The optimised parameters follow a Gaussian distribution.  The optimised parameters resulting from the search are also given in Table~\ref{noise_1} and Table~\ref{noise_2}; these parameters are close to the values given in FS18.  The optimised parameters are typically the same as the input guess apart from the fact that the rms values are about 1.7 to 2 times smaller than the input rms values.  Figure~\ref{correls} gives two typical examples optimised correlation to compare with Fig. 1 and Fig. 2 in FS18; the two typical examples are the median and the maximum of the sum of the correlations (obtained in Step 3 above).

\begin{table*}[t]
\caption{Guess parameters and parameters optimising the correlation for $l=1$ for a white Gaussian noise.  The first column gives the asymptotic period, the second column gives the minimal period, the third column gives the deviation from the asymptotic formula; the fourth column gives the splitting and the fifth column gives the amplitude ratio of $m=1$ to $m=0$.}             
\label{fitted}      
\centering                          
\begin{tabular}{cccccc}        
\hline                 
Case&$P_0$ (in s) &$P_{\rm min,1}$ (in s)&$\alpha_1$ (in s)&$\nu_1$ (in $\mu$Hz)&$w_{1,1}$\\
\hline
Guess&2042 $\pm$ 30&33252 $\pm$ 1000&1339 $\pm$ 30&0.209 $\pm$ 0.05&0.60 $\pm$ 0.40\\
Optimised&2041 $\pm$ 17&33250 $\pm$ 560&1339 $\pm$ 20&0.209 $\pm$ 0.03&0.60 $\pm$ 0.20\\ 
\hline  
\end{tabular}
\label{noise_1}
\end{table*}

\begin{table*}[t]
\caption{Guess parameters and parameters optimising the correlation for $l=2$ for a white Gaussian noise.  The first column gives the asymptotic period, the second column gives the minimal period, the third column gives the deviation from the asymptotic formula; the fourth columns gives the splitting, the last two column give the amplitude ratio of $m=1$ to $m=0$, and of $m=2$ to $m=0$.}             
\label{fitted}      
\centering                          
\begin{tabular}{ccccccc}        
\hline                 
Case&$P_0$ (in s) &$P_{\rm min,2}$ (in s)&$\alpha_2$ (in s)&$\nu_2$ (in $\mu$Hz)&$w_{2,1}$&$w_{2,2}$\\
\hline
Guess&2040 $\pm$ 30&31884 $\pm$ 1000&1319 $\pm$ 30&0.628 $\pm$ 0.05&0.60 $\pm$ 0.40&0.30 $\pm$0.40\\
Optimised&2040 $\pm$ 17&31880 $\pm$ 570&1320 $\pm$ 20&0.629 $\pm$ 0.03&0.65 $\pm$ 0.20&0.34 $\pm$ 0.20\\ 
\hline
\end{tabular}
\label{noise_2}
\end{table*}

\begin{figure*}[!]
\centerline{
\vbox{
\hbox{
\includegraphics[width=6 cm,angle=90]{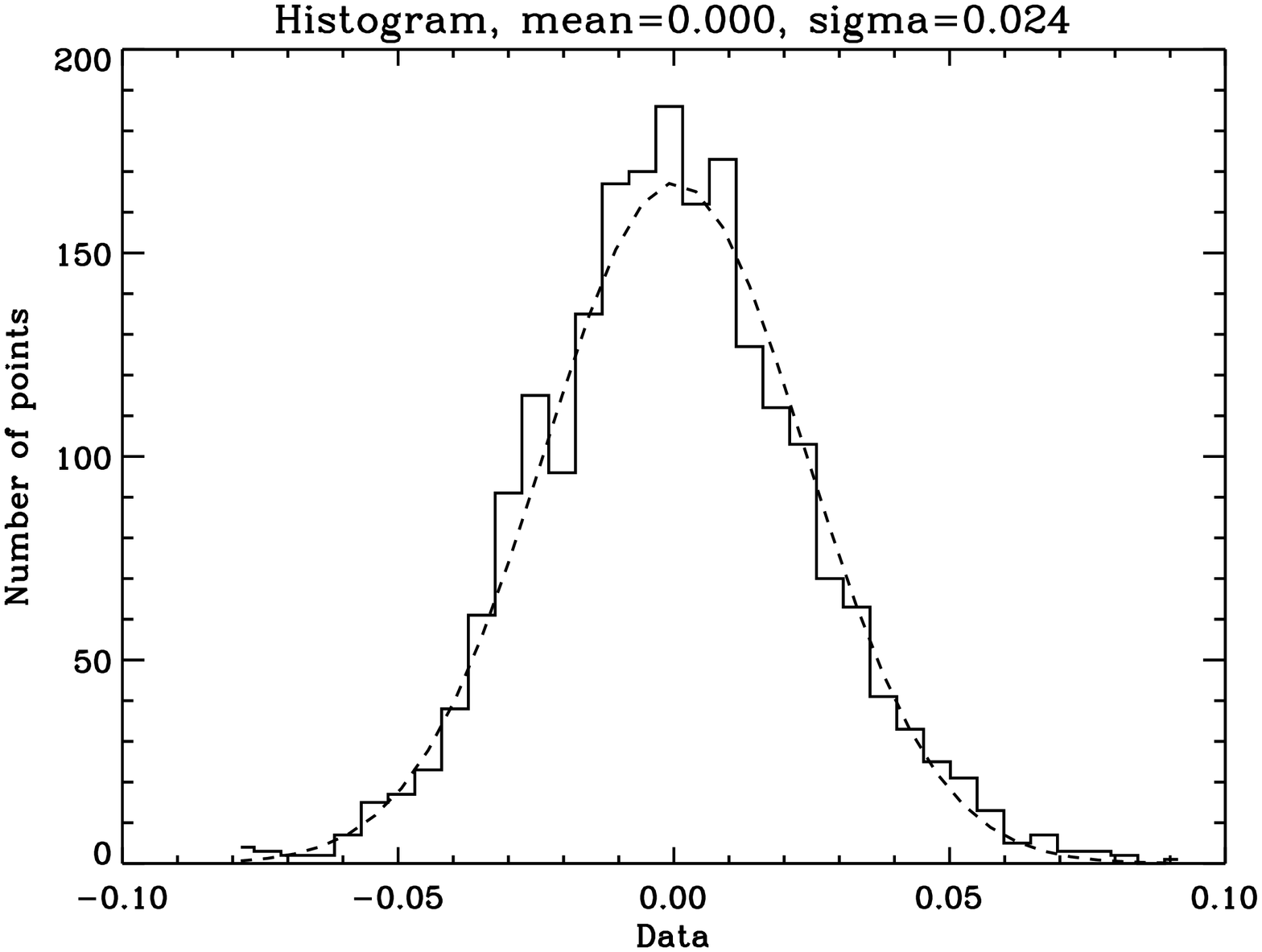}
\includegraphics[width=6 cm,angle=90]{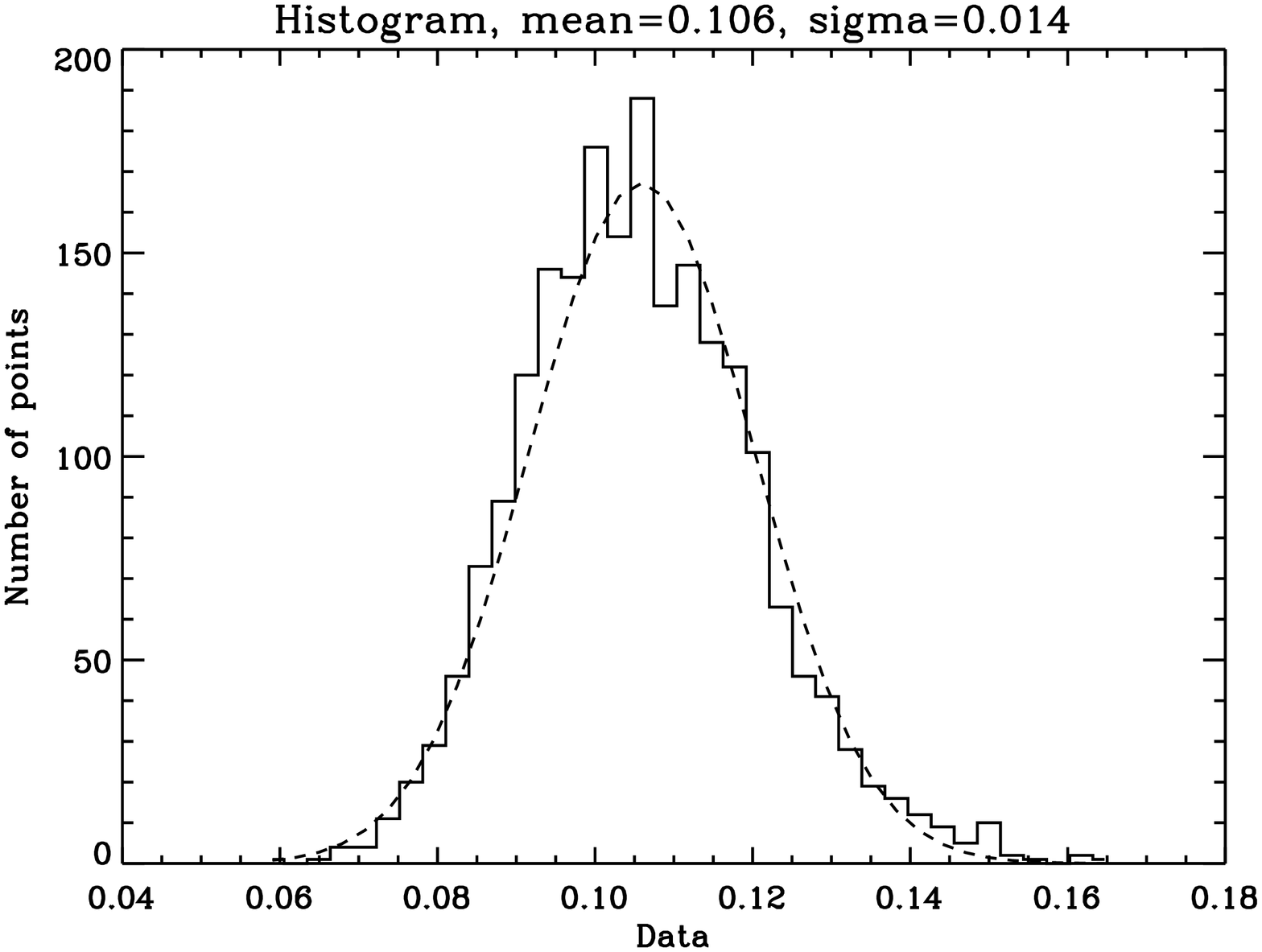}}
}}
\caption{Histogram of the correlation at lag=0 {before} optimisation (left) and {after} optimisation (right) for $l=1$.
}
\label{histo}
\end{figure*}

Figure~\ref{histo} provides the measured distribution of the correlation for lag=0 before and after optimisation for $l=1$.  We can also derive from the simulation the distribution of the correlation for lag$\ne$0 which is a Gaussian with a mean value of 0, and rms of 0.017.  In FS18, they use this rms value as a measurement of the rms value of the fluctuation of the correlation at lag=0, then they gave a signal-to-noise ratio of about 8 for lag=0.  As a matter of fact, as shown by Figure~\ref{histo}, due to the optimisation, the signal-to-noise ratio is bound to be on average 6.2 when using the 0.017 figure (6.2=0.106/0.017); and 10\% of the simulation gives a value higher than 7.3 $\sigma$.  In addition, the rms value used in FS18 is derived for lag$\ne$0 while the rms value at lag=0 is higher as shown in Figure~\ref{histo} (0.024 instead of 0.017); such that our figures for the average and the 10\% percentile should be lower by 1.4.  The cross-correlation between the model spectrum and the observed (or simulated) power spectrum being the Fourier transform of the cross power spectrum, it is expected that the variance at lag=0 should be twice the variance at lag$\ne$0. Of course, our figures cannot be directly compared with FS18 because the noise simulated here under the null hypothesis is a white noise, which is not the case of the true data.  Nevertheless, we can draw three conclusions from these simulations:
\begin{itemize}
\item White noise can provide correlation similar to Fig. 1 and Fig. 2 in FS18 (contradicting the findings of that paper)
\item The statistics of the correlation at lag=0 after optimisation has different mean and rms value than those of the statistics of the correlation at lag$\ne$0
\item The statistics of the correlation at lag=0 can not be compared to the statistics of the correlation at lag$\ne$0
\end{itemize}




\subsection{GOLF signal}
\subsubsection{Same parameters as in FS18}
\begin{figure}[!]
\centerline{
\includegraphics[width=7 cm,angle=90]{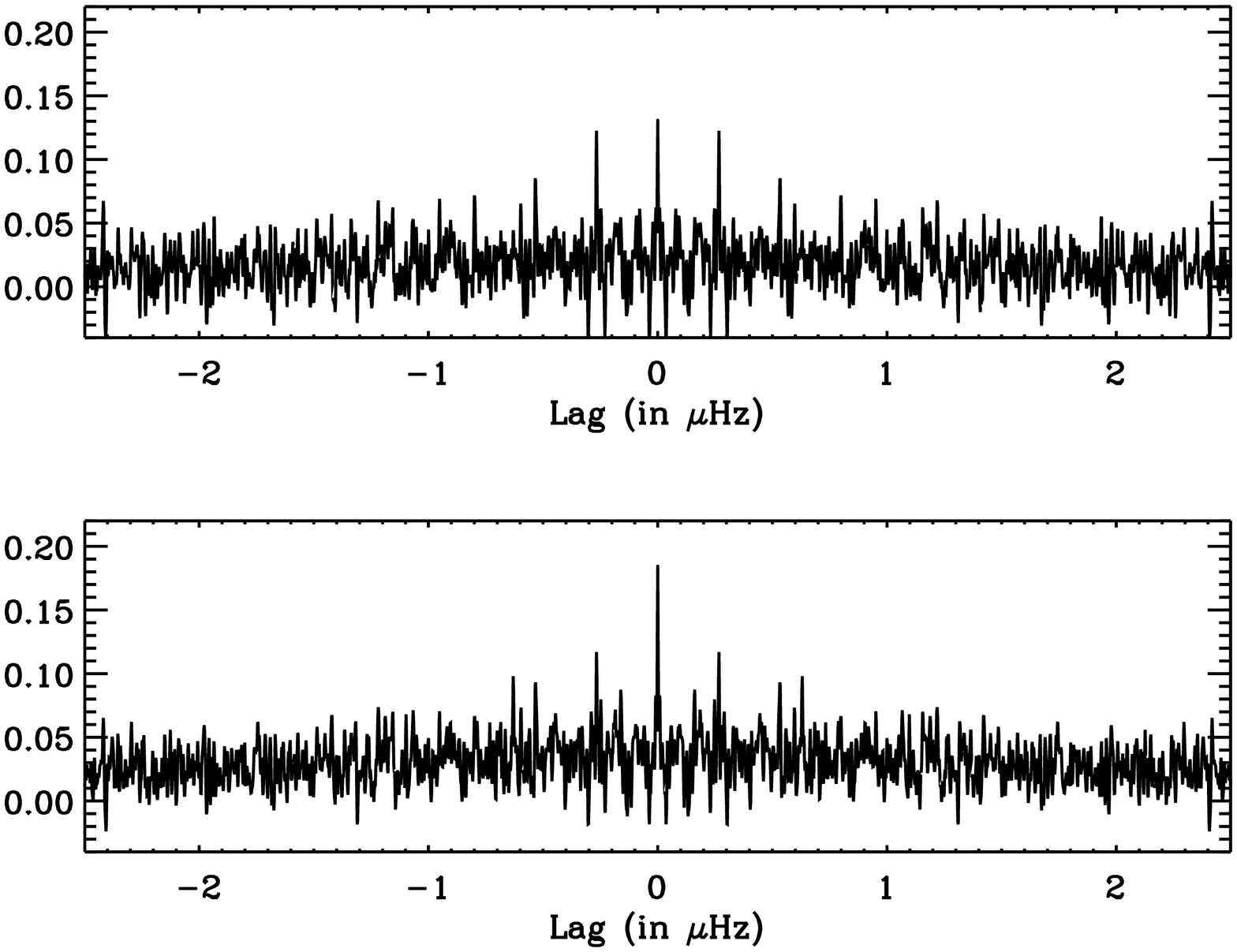}
}
\caption{{Same set of parameters as FS18}.  Correlation of the g-mode model spectra with the power spectrum of round-trip travel time for $l=1$ (Top; similar to Fig. 1 in FS18) and for $l=1$ and $l=2$ (Bottom; similar to Fig. 2 in FS18)
}
\label{optim_1}
\end{figure}

We then used the optimisation procedure above for finding the best set of parameters for the GOLF data, starting with parameter values similar to those of the FS18.  The results are shown in Fig.~\ref{optim_1}.  We show that we can reproduce the results of FS18 that they obtained in their figures 1 and 2.

\subsubsection{Different parameters from FS18}
We also applied the optimisation procedure to three different initial guesses of parameter values.  The parameters differed from the starting set of FS18 as follows:
\begin{itemize}
\item $P_{\rm min,1}$ increased by 721 s, and $P_{\rm min,2}$ increased by 416 s
\item $P_0$ increased by 25\%.
\item $P_0$ increased by 50\%.
\end{itemize}
These parameters were chosen such that the location of the g-mode periods are completely independent from the periods obtained by FS18.  The first set is rather drastic since it shifts all g-mode periods by a few hundreds seconds.  For orders close to $n_0$, it simply shifts the mode period in between the $n_0$ and the $n_0+1$
of the parameters set of FS18.  The two other sets are just pertinent for different kinds of stellar models, even though these $P_0$ values are unrealistic for a solar model.  These different values of $P_0$ are unrealistic for a solar model as typical ranges are about a few \% from the solar standard model.

\begin{figure}[!]
\centerline{
\includegraphics[width=7 cm,angle=90]{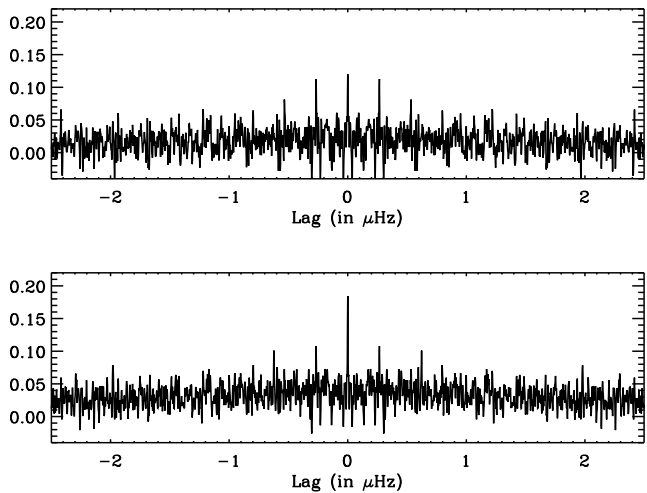}
}
\caption{{Different $P_{\rm min,l}$ from FS18}.  Correlation of the g-mode model spectra with the power spectrum of round-trip travel time for $l=1$ (Top; similar to Fig. 1 in FS18) and for $l=1$ and $l=2$ (Bottom; similar to Fig. 2 in FS18)
}
\label{optim_5}
\end{figure}

\begin{figure}[!]
\centerline{
\includegraphics[width=7 cm,angle=90]{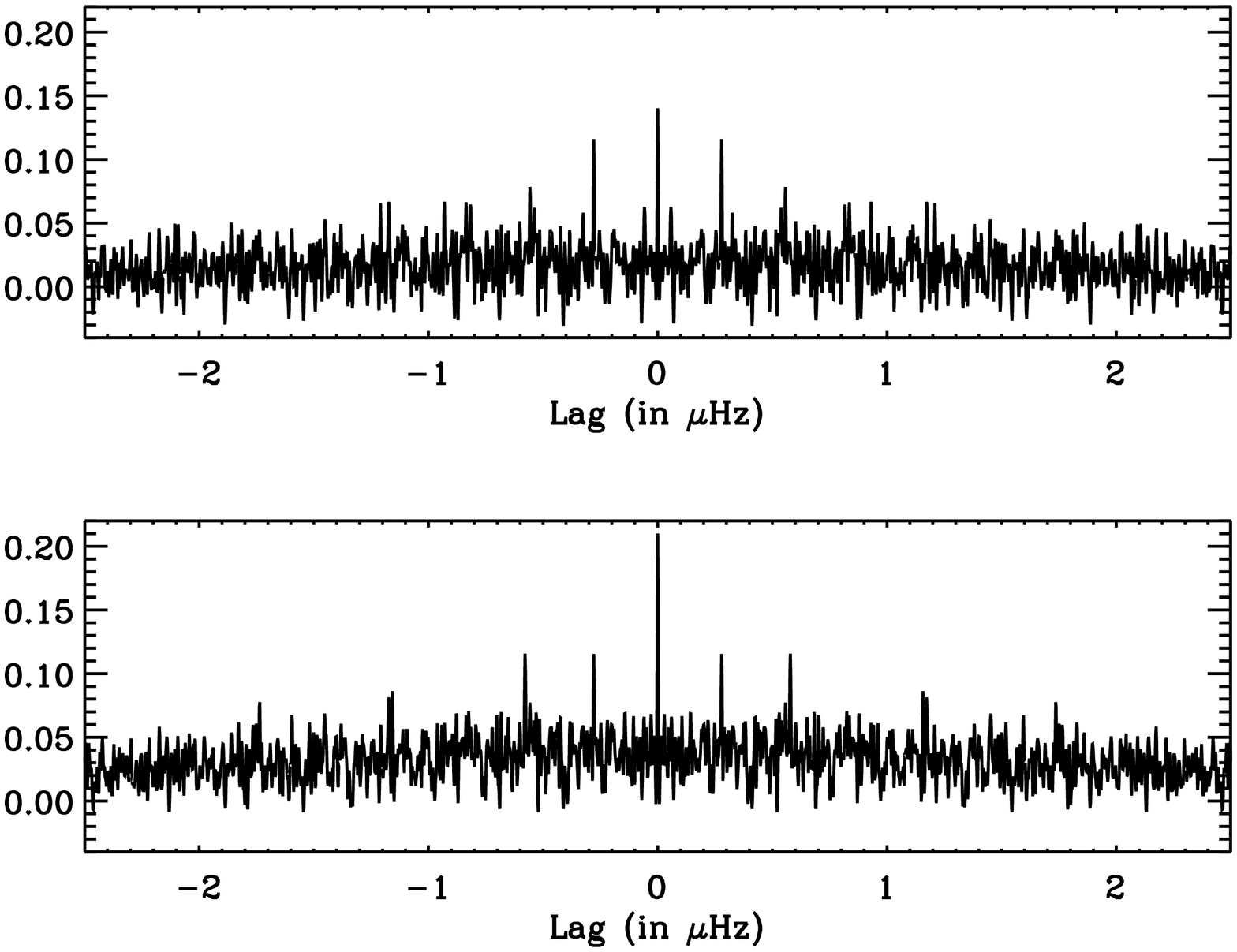}
}
\caption{{$P_0$ increased by 25\% from FS18}.  Correlation of the g-mode model spectra with the power spectrum of the time series of round-trip travel time for $l=1$ (Top; similar to Fig. 1 in FS18) and for $l=1$ and $l=2$ (Bottom; similar to Fig. 2 in FS18)
}
\label{optim_6}
\end{figure}

\begin{figure}[!]
\centerline{
\includegraphics[width=7 cm,angle=90]{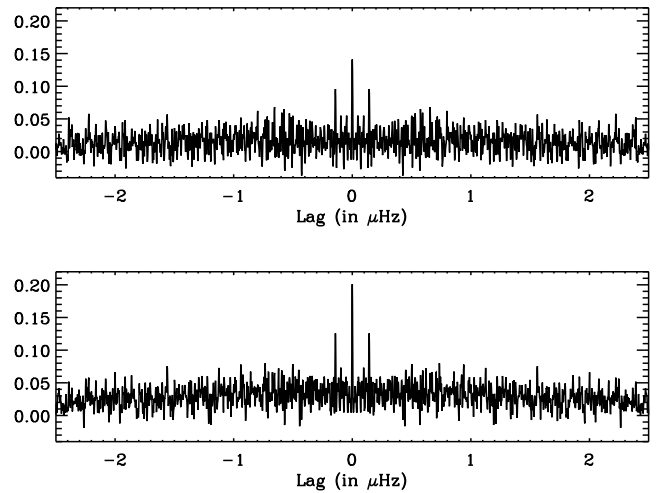}
}
\caption{{$P_0$ increased by 50\% from FS18}.  Correlation of the g-mode model spectra with the power spectrum of the time series of round-trip travel time for $l=1$ (Top; similar to Fig. 1 in FS18) and for $l=1$ and $l=2$ (Bottom; similar to Fig. 2 in FS18)
}
\label{optim_7}
\end{figure}

The results are shown in Figs.~\ref{optim_5}, \ref{optim_6} and \ref{optim_7}.  We can clearly see high correlation at lag=0 and at lag=$\nu_s$ but for very different values of $P_0$ and different values of $P_{\rm min,1}$ and $P_{\rm min,2}$ similar to those of Fig.~\ref{optim_1} and Fig.~\ref{correls}.

\begin{table*}[t]
\caption{Guess parameters and parameters optimising the correlation for for $l=1$ for 4 different cases applied to the GOLF data.  The first column gives the asymptotic period, the second column gives the minimal period, the third column gives the deviation from the asymptotic formula; the fourth columns gives the splitting and the fifth column gives the amplitude ratio of $m=1$ to $m=0$.  The parameters differing from F17 and II are shown in bold.}             
\label{fitted}      
\centering                          
\begin{tabular}{cccccc}        
\hline                 
Case&$P_0$ (in s) &$P_{\rm min,1}$ (in s)&$\alpha_1$ (in s)&$\nu_1$ (in $\mu$Hz)&$w_{1,1}$\\
\hline  
\multicolumn{6}{c}{} \\
\multicolumn{6}{c}{Same set as FS18} \\
\hline
Guess&2042 $\pm$ 30&33252 $\pm$ 1000&1339 $\pm$ 30&0.209 $\pm$ 0.05&0.60 $\pm$ 0.40\\
Best&2028.05&32979.85&1341.56& 0.267 &0.736\\
\hline  
\multicolumn{6}{c}{} \\
\multicolumn{6}{c}{$P_{\rm min,1}$ different from FS18} \\
\hline
Guess&2042 $\pm$ 30&{\bf 33973 $\pm$ 1000}&1339 $\pm$ 30&0.209 $\pm$ 0.05&0.60 $\pm$ 0.40\\
Best&2028.10&{\bf 32982.67}&1311.48&0.267&0.703\\
\hline  
\multicolumn{6}{c}{} \\
\multicolumn{6}{c}{$P_{0}$ increased by 25\% from FS18} \\
\hline
Guess&{\bf 2552 $\pm$ 30}&33252 $\pm$ 1000&1339 $\pm$ 30&0.209 $\pm$ 0.05&0.60 $\pm$ 0.40\\
Best&{\bf 2567.39}   &    32562.60     &  1346.59  &    0.278  &    0.625\\
\hline  
\multicolumn{6}{c}{} \\
\multicolumn{6}{c}{$P_{0}$ increased by 50\% from FS18} \\
\hline
Guess&{\bf 3063 $\pm$ 30}&33252 $\pm$ 1000&1339 $\pm$ 30&0.209 $\pm$ 0.05&0.60 $\pm$ 0.40\\
Best&{\bf 3056.68}     &  32560.09    &   1355.11   &   0.142    &  0.692\\
\hline
\end{tabular}
\label{golf_1}
\end{table*}

\begin{table*}[t]
\caption{Guess parameters and parameters optimising the correlation for $l=2$ for 4 different cases applied to the GOLF data.  The first column gives the asymptotic period, the second column gives the minimal period, the third column gives the deviation from the asymptotic formula; the fourth columns gives the splitting, and the last two columns give the amplitude ratio of $m=1$ to $m=0$, and of $m=2$ to $m=0$.  The parameters differing from F17 and II are shown in bold.}             
\label{fitted}      
\centering                          
\begin{tabular}{ccccccc}        
\hline                 
Case&$P_0$ (in s) &$P_{\rm min,2}$ (in s)&$\alpha_2$ (in s)&$\nu_2$ (in $\mu$Hz)&$w_{2,1}$&$w_{2,2}$\\
\hline  
\multicolumn{7}{c}{} \\
\multicolumn{7}{c}{Same set as FS18} \\
\hline
Guess&2040 $\pm$ 30&31884 $\pm$ 1000&1319 $\pm$ 30&0.628 $\pm$ 0.05&0.60 $\pm$ 0.40&0.30 $\pm$0.40\\
Best &2039.45    &   31901.31    &   1339.39   &   0.630   &   0.655   &   0.528 \\
\hline  
\multicolumn{7}{c}{} \\
\multicolumn{7}{c}{$P_{\rm min,2}$ different from as FS18} \\
\hline
Guess&2040 $\pm$ 30&{\bf 32300 $\pm$ 1000}&1319 $\pm$ 30&0.628 $\pm$ 0.05&0.60 $\pm$ 0.40&0.30 $\pm$0.40\\
Best&2043.86 &      {\bf 32657.81}   &    1316.32 &     0.622     & 0.890   &   0.568\\
\hline  
\multicolumn{7}{c}{} \\
\multicolumn{7}{c}{$P_{0}$ increased by 25\% from FS18} \\
\hline
Guess&{\bf 2549 $\pm$ 30}&31884 $\pm$ 1000&1319 $\pm$ 30&0.628 $\pm$ 0.05&0.60 $\pm$ 0.40&0.30 $\pm$0.40\\
Best&{\bf 2528.79}    &   32359.56    &   1294.83 &     0.578   &   0.749   &   0.516\\
\hline  
\multicolumn{7}{c}{} \\
\multicolumn{7}{c}{$P_{0}$ increased by 50\% from FS18} \\
\hline
Guess&{\bf 3060 $\pm$ 30}&31884 $\pm$ 1000&1319 $\pm$ 30&0.628 $\pm$ 0.05&0.60 $\pm$ 0.40&0.30 $\pm$0.40\\
Best&{\bf 3058.91}   &    31614.12  &     1307.81  &    0.621    &  0.707    &  0.480\\
\hline
\end{tabular}
\label{golf_2}
\end{table*}

\subsection{Summary}
Here we give a short summary of our study of the g-mode detection claims of FS18:
\begin{itemize}
\item Proper optimisation of the parameters is key for obtaining high correlation values at a given lag.
\item Correlation mimicking g-mode detection is obtained with pure white Gaussian noise for $l=1$ and $l=2$
\item Correlation mimicking g-mode detection is obtained with the GOLF data with a very different set of parameters ($P_{\rm min,l}$, $P_0$) for $l=1$ and $l=2$
\end{itemize}
There were no tests performed for the g-mode detection claim for $l=3$ and $l=4$.  This is discussed in the next section.

\section{Discussion}
In Section 2, we confirmed the {\rm fragility}, as coined by \citet{Schunker2018}, of the g-mode detection in F17.  Different photomultipliers, shorter sub-series and to some extent sampling different from 80 s do not support the g-mode detection of F17.  In addition, the change of the original cadence of 4 hours and the shifting of the start time provides the same lack of reproduction \citep{Schunker2018}.  The impact of changes in the sampling (less than 80 s), the changes in the cadence (3h to 5h), and the start time (shifted by less than 2 hours) are all in a time domain that should not affect the detection of the g modes having periodicity ranging from 8 h to 40 h.  The 20-s sampling affects much less the folding of high frequencies above the Nyquist frequency than for the 60-s and 80-s samplings.  The detection is also not confirmed for the 60-s sampling while there is an indication of detection for the other 2 samplings.   In addition, the fact that the detection of F17 is not confirmed when shifting the start of the time series or with different photomultipliers is an indication that we might suspect noise noise mimicking what at first look appears to be
significant signal.  Would the signal had been really present in GOLF, it would have showed up irrespective of the sampling, the cadence or the photomultipliers.  Last but not least, the detection of F17 could not be confirmed with other instruments such as BiSON or GONG.

We must also point out two problems with the asymptotic formula used in F17 and FS18.   First, Eq. (5) gives the formula used in F17 for which $\alpha$ can also be derived from the asymptotic formula of \citet{JPGB1986} as $\alpha=P_l(l(l+1)V_1+V2)$.  Using the values given in \citet{JPGB1986} for $V_1$ and $V_2$, we can derive for $l=1$ and $l=2$, $\alpha_1$=9466 s, and $\alpha_2$=687~s.  These two values are about a factor 5 to 7 larger than those inferred in F17 (for $l=1$) and in FS18 (for $l=1$ and $l=2$).  Since $\alpha$ is directly related to $P_0$, there is an inconsistency in getting two $P_l$ (for $l=1$ and $l=2$) consistent with a common $P_0$, while having the $\alpha$'s not consistent with $P_0$.  Of course, there could be a solar model leading to $V_1$ and $V_2$ values able to match such a low value of $\alpha$ , but this model remains to be found.  Second, the asymptotic formula used in F17 is the same as Eq. (5).  Unfortunately, Eq. (5) is an approximation of Eq. (1) that differs by ${\rm O}(1/n^3)$.  The resulting difference in  mode frequency is larger than several resolution bins for dozens of low order modes both for $l=1$ and $l=2$.  As a matter of fact, the difference can be somehow compensated by the factor $\alpha/n$ but at the expense of introducing a bias to $\alpha$.  This second defect may not affect that much the detection since only about 10\% of the modes are affected by this inaccuracy.  Of course, we note that the asymptotic formula derived under the Cowling approximation and with $\theta$ constant  may itself provide some errors when compared to a solar model \citep{JPGB1986}.

In Section 3, we could not confirm the detection of g modes as reported in FS18, nor could we confirm the fact that noise cannot produce such correlation.  Rather, it is clear that noise can reproduce the correlation found in FS18 for $l=1$ and $l=2$, but in addition any guess parameter can also reproduce high correlation (even higher correlation) using the GOLF data.  In short, given
the hundreds of peaks available in the model spectrum for $l=1$ and $l=2$, it is always possible to find a high correlation.  As a matter of fact, this is the very reason why the correlation also permits the putative detection of higher degrees ($l > 2$).  This fact is simply an artefact of the detection methodology.  

The maximisation of the correlation with the model spectrum involves the coincidence of several hundreds of peaks being at the same location in the observed (or simulated) spectrum.  When trying to optimise the correlation, one tries to optimise the average level of the ensemble of the hundreds of peaks; here the optimisation tries to maximise the average level of these peaks.   Therefore, we can explain this artefact by using a simple example not for many peaks but just a single peak.  The optimisation for the single peak can be put in parallel with the statistics of the maximum.  For example assuming the same probability distribution function (pdf) of an ensemble of $n$ i.i.d.(independent and identically distributed) random variable $X$, what is the statistics of the maximum $Y$ of these $n$ random variables?  The solution is given by computing the cumulative distribution function (cdf) as:
\begin{equation}
P(Y<y)=P(X_1<y;...;X_n<y)=[P(X<y)]^n
\end{equation}
then the pdf of $Y$ ($p_Y$) is simply given as the derivative of the cdf with respect to $y$ as
\begin{equation}
p_Y(y)=n p_X(y) [P(X<y)]^{(n-1)}
\end{equation}
where $p_X$ is the cdf of $X$.  It is obvious that the pdf of $X$ and $Y$ are different since we have:
\begin{equation}
p_Y(y) \ne  p_X(y)
\end{equation}
This example is used for a single ensemble of $X$ random variables.  In the case of the optimisation of the correlation, we find the maximum amongst not a single ensemble but amongst several hundreds of ensembles related to the number of g-mode orders and degree.  The number $n$ of variables for a single ensemble (i.e. a single g-mode peak) depends upon how we set the size of the hypersphere and the resolution in the power spectrum.  In theory, the resulting pdf of the sum of the maximised peaks can be computed but this is rather complicated as some modes of different orders overlap.  In practice, it is somewhat easier to do the Monte-Carlo simulation that we performed in Section 3.  Even though we do not write an analytical formulation for the correlation, we understand well the reason why the correlation at lag=0 can be higher by several $\sigma$ compared to the basal value of the correlation.  Of course the explanation for lag=$\nu_l$ is the same since we also optimise the correlation at that lag.

In light of this work we would like to stress that an important discovery such as detecting solar g-modes should probably rely on a higher detection level than used before.  For instance, the 10\% probability level provides a posterior probability of H$_{0}$ to be true of at least 38\%.  Here we suggest to adopt the lower detection level of 1\%, providing a posterior probability H$_{0}$ to be true of at least 11\%.  As already stressed by \citet{Appourchaux2011}, this conservative approach is the result of a Bayesian framework.  We also suggest a higher rejection level of 1\% instead of the canonical 10\% as already suggested by \citet{Appourchaux2010}.

\section{Conclusion}
We cannot confirm the g-mode detection claims of F17 and FS18.  The lack of reproducibility of the detection in F17 is most likely due to the noise contributing in different manners depending on how the GOLF data are analysed (sampling, cadence, photomultiplier, start time).  Even though the detection is not confirmed, F17 has restarted the work on g-mode detection, which is greatly  beneficial to the field. We believe that the detection of g modes as made in FS18 is the result of an artefact of the analysis methodology related to the large number of modes used in the analysis.  The artefact also explains the claimed detection of g modes of degree higher than 2.


\begin{acknowledgements}
The GOLF instrument on board SoHO is based upon a consortium of institutes (IAS, CEA/Saclay, Nice and Bordeaux
Observatories from France, and IAC from Spain) involving a cooperative effort of scientists, engineers, and technicians, to whom we are indebted.  GOLF data are available at the MEDOC data and operations centre (CNES / CNRS /Univ. Paris-Sud) at medoc.ias.u-psud.fr.
SoHO is a mission of international collaboration between ESA and NASA.  This work utilises data obtained by the Global Oscillation Network Group (\,GONG\,) program, managed by the National Solar Observatory, which is operated by AURA, Inc., under a cooperative agreement with the National Science Foundation. The data were acquired by instruments operated by the Big Bear Solar Observatory, High Altitude Observatory, Learmonth Solar Observatory, Udaipur Solar Observatory, Instituto de Astrof\'{\i}sica de Canarias, and Cerro Tololo Interamerican Observatory.   We are grateful to Patrick Boumier for many interesting discussions.  We are thankful to l'Amerloque for proof reading and critical discussions.  This paper is dedicated to the everlasting memory of our friend and colleague, Claus Fr\"ohlich, who was always ready to search for solar g modes; your legacy shall last.
\end{acknowledgements}

\bibliographystyle{aa}
\bibliography{thierrya}

\appendix

\end{document}